\documentclass[showpacs,aps,prd,reprint,superscriptaddress,nofootinbib,longbibliography,preprintnumbers]{revtex4-1}
\usepackage[colorlinks=true, pdfstartview=FitV, linkcolor=magenta,citecolor=blue, urlcolor=magenta,
bookmarks=true, bookmarksnumbered=true, breaklinks]{hyperref}
\usepackage[dvipdfmx]{graphicx}

\usepackage{here}
\usepackage{url}
\usepackage{amsmath,amssymb,bm,color,longtable,mathrsfs,amsfonts,slashed}

\newcommand{\Slash}[1]{{\ooalign{\hfil/\hfil\crcr$#1$}}}

\begin{document}
\begin{flushright}
\end{flushright}

\title{Roles of $U(1)$ axial anomaly effects in cold and dense two-color QCD with $2+2$ flavors}

\author{Manato Sakai}
\affiliation{Department of Physics, Nagoya University, Nagoya 464-8602, Japan}
\author{Daiki Suenaga}
\affiliation{Kobayashi-Maskawa Institute for the Origin of Particles and the Universe, Nagoya University, Nagoya 464-8602, Japan}
\affiliation{Research Center for Nuclear Physics, Osaka University, Ibaraki 567-0048, Japan}

\date{\today}

\begin{abstract}
We explore phase structures and hadron mass spectra in cold and dense two-color QCD with $2+2$ flavors where the sign problem disappears. We particularly focus on $U(1)$ axial anomaly effects. We employ an $N_f=2+2$ linear sigma model based on the $SU(8)$ Pauli-G\"{u}rsey symmetry to describe negative-parity as well as positive-parity hadrons, for which low-energy excitations are appropriately described particularly in the baryon superfluid phase with light diquark condensates. As a result, the strange chiral condensate is found to be enhanced in the superfluid phase owing to the flavor-mixing structure of $U(1)$ axial anomaly effects. We also confirm this enhancement by means of the Nambu--Jona-Lasinio model. Besides, our present analysis predicts the existence of a novel superfluid phase where heavy diquarks condense in dense regime. The topological susceptibility with $2+2$ flavors from the viewpoints of the anomaly effects and chiral-symmetry restoration is investigated. Furthermore, we derive a complete inverse mass hierarchy for negative-parity diquarks with sufficient anomaly effects. Our findings are expected to provide future lattice simulations with useful information on flavor-symmetry violation from the $U(1)$ axial anomaly aspects in cold and dense two-color QCD medium.
 \end{abstract}


\maketitle

\section{Introduction}
\label{sec:Introduction}


Quantum chromodynamics (QCD) in cold and dense system is expected to exhibit various aspects in the presence of $s$ quarks with the violation of $SU(3)$ flavor symmetry. At hadronic level, the so-called {\it hyperon puzzle} concerning appearances of strange baryons in medium, in the context of the stiffness of neutron-star equation of state, is known as a significant issue~\cite{Baym:2017whm}. At quark level, meanwhile, the color superconducting phase is predicted to undergo the {\it two-flavor superconductivity} (2SC) where quark Cooper parings are formed by only $u$ and $d$ quarks, before reaching the {\it color-flavor locking} (CFL) phase~\cite{Alford:2007xm}. In order to gain more insights into such phenomena, deeper understandings of flavor-symmetry violation together with chiral-symmetry restoration are essential.

Despite abundant theoretical predictions, there are no handleable experiments to directly confirm structures in cold and dense QCD. Instead, it is worth focusing on QCD-like theories such as two-color (QC$_2$D). In QC$_2$D with even numbers of flavors, the {\it sign problem} of Monte-Carlo simulations with baryon chemical potential disappears, and thus lattice QCD simulations are straightforwardly applicable~\cite{Aarts:2015tyj,Nagata:2021ugx}. In other words, QC$_2$D allows us to conduct {\it numerical experiments} of cold-dense system. Thus far, many examinations in QC$_2$D have been done from both lattice simulations~\cite{Muroya:2003qs,Boz:2019enj,Buividovich:2020dks,Astrakhantsev:2020tdl,Iida:2024irv,Braguta:2023yhd} and model approaches~\cite{Kogut:2000ek,Lenaghan:2001sd,Sun:2007fc,Brauner:2009gu,Harada:2010vy,Strodthoff:2011tz,Imai:2012hr,Duarte:2015ppa,Contant:2019lwf,Suenaga:2019jjv,Khunjua:2020xws,Suenaga:2022uqn,Acharyya:2024pqj}.

On the QC$_2$D lattice, flavor-symmetry violating system at finite density is realized by introducing heavy quarks $(Q_1,Q_2)$ as well as light ones $(q_1,q_2)$, where each sector shares the identical mass so as to cancel out the sign problem. In the present paper, we construct a linear sigma model (LSM) with $2+2$ flavors based on the Pauli-G\"{u}rsey $SU(8)$ symmetry~\cite{Kogut:1999iv,Kogut:2000ek}, as an extension of the $N_f=2$ version which has been developed in the literatures~\cite{Suenaga:2019jjv,Kawaguchi:2023olk,Suenaga:2023xwa,Kawaguchi:2024iaw,Fejos:2025nvd,Fejos:2025oxi,Suenaga:2025sln}, in order to provide future lattice simulations with useful information on flavor-symmetry violation in cold-dense medium.

In QC$_2$D, diquarks become color-singlet hadrons so that they form Bose-Einstein condensates (BECs) when the quark chemical potential exceeds a critical value. In $N_f=2+2$ case the light diquarks ($\sim qq$) first start to create the BECs. This phase is referred to as the (light) {\it diquark condensed phase}, or (light) {\it baryon superfluid phase} due to color-singlet nature of the diquarks. Meanwhile, the stable phase with no diquark condensates is often called {\it hadronic phase}. The baryon superfluid phase has been indeed predicted by effective models and confirmed by lattice simulations (see, e.g., Ref.~\cite{Suenaga:2025sln} and references therein).

In the present paper, in particular we study the fate of the heavy chiral condensate $\langle\bar{Q}Q\rangle$ and the mass spectrum of hadrons containing $Q$ ($\bar{Q}$). We pay much attention to the $U(1)$ axial anomaly effects, which are known to play an important role in crossover transition from hadronic matter to quark matter in three-color QCD~\cite{Schafer:1998ef,Abuki:2010jq}. The anomaly induces mixings of the flavors in the effective-model treatment, as represented by the well-known Kobayashi-Maskawa-'t Hooft (KMT) determinant interaction~\cite{Kobayashi:1970ji,Kobayashi:1971qz,tHooft:1976snw,tHooft:1976rip}. Hence, in the superfluid phase where only the light diquarks are condensed, we can expect distinctive phenomena for the heavy sector driven by the anomaly effect. 
We also investigate the topological susceptibility which is strongly related to the $U(1)$ axial anomaly to see roles of the heavy flavors~\cite{Kawaguchi:2023olk,Iida:2024irv,Braguta:2023yhd}.

Examination of diquarks in QC$_2$D allows us to gain deeper insights into their own properties thanks to the color-singlet nature, whereas in our three-color QCD world the colorful diquark dynamics can be implicitly inspected through singly heavy baryons (SHBs)~\cite{Harada:2019udr}. For this advantage, in the present study we also focus on diquark masses from a viewpoint of the flavor-mixing structures of the $U(1)$ axial anomaly effects, apart from investigations in cold and dense medium.

This paper is organized as follows. In Sec.~\ref{sec:Model}, we briefly explain disappearance of the sign problem in QC$_2$D with $N_f=2+2$, and then introduce our $N_f=2+2$ LSM. In Sec.~\ref{sec:Numerical} we present numerical results on the mean fields, hadron masses and topological susceptibility in cold and dense QC$_2$D medium with $2+2$ flavors. In Sec.~\ref{sec:NJLAnalysis} a simple analysis based on the Nambu--Jona-Lasinio (NJL) model is given, and Sec.~\ref{sec:InverseMH} is devoted to examination of the diquark mass hierarchy with the $U(1)$ axial anomaly. Finally, in Sec.~\ref{sec:Conclusions} we conclude the present study.

\section{Model}
\label{sec:Model}

\subsection{Properties of $N_f=2+2$ QC$_2$D Lagrangian}
\label{sec:QC2DLagrangian}

Before constructing the LSM describing low-lying hadrons in $N_f=2+2$ QC$_2$D, here we briefly show fundamental properties of QC$_2$D Lagrangian.

First, we explain disappearance of the sign problem in $N_f=2+2$ QC$_2$D. In the Euclidean notation, the QC$_2$D Lagrangian with two light flavors $(q_1,q_2)$ and two heavy flavors $(Q_1,Q_2)$ takes the form of
\begin{eqnarray}
{\cal L}_{\rm QC_2D}^{N_f=2+2} = \bar{\psi}(\gamma\cdot {D} -\gamma_4\hat{\mu}+\hat{M})\psi\ , \label{QC2DLagrangian}
\end{eqnarray}
where $\psi = (q_1,q_2,Q_1,Q_2)^T$ is a quark quartet. The covariant derivative ${D}_\mu\psi = (\partial_\mu-ig_s A^a_\mu \tau_c^a/2)\psi$ incorporates interactions among quarks $\psi$ and gluons $A^a_\mu$, with a strong coupling constant $g_s$ and the Pauli matrix $\tau_c^a$. The remaining matrices $\hat{\mu}$ and $\hat{M}$ represent the chemical potential and quark masses, respectively. In the present analysis we assume ($m_q<m_Q$)
\begin{eqnarray}
\hat{\mu} &=& \mu_q \hat{I}_q  + \mu_Q\hat{I}_Q \ ,\nonumber\\
\hat{M} &=& m_q \hat{I}_q  + m_Q\hat{I}_Q\ , \label{MuAndM}
\end{eqnarray}
where 
\begin{eqnarray}
\hat{I}_q = {\rm diag}(1,1,0,0)\ , \ \ \hat{I}_Q = {\rm diag}(0,0,1,1)\ .
\end{eqnarray}
We note that Euclidean gamma matrices are defined so as to satisfy the anti-commutation relation $\{\gamma_\mu,\gamma_\nu\} = 2\delta_{\mu\nu}$ ($\mu=1$ - $4$)  and Hermicity $\gamma_\mu = \gamma_\mu^\dagger$.

Since off-diagonal components connecting light and heavy flavors in the four-flavor Dirac operator ${\cal D} \equiv \gamma\cdot {D} + \hat{M}$ are absent, path integrals with respect to $\psi$ and $\bar{\psi}$ yield a simple product of
\begin{eqnarray}
\prod_{f=q,Q}\Big[{\rm Det}(\gamma\cdot D-\gamma_4\mu_f+m_f) \Big]^2\ , \label{DetProd}
\end{eqnarray}
where the symbol ``Det'' stands for determinants with respect to Euclidean space, colors and Dirac indices. The square stems from the degeneracies of $q_1$ and $q_2$ or $Q_1$ and $Q_2$. Here, one can easily show that ($D_\mu^\dagger = -D_\mu$)
\begin{eqnarray}
&& {\rm Det}(\gamma\cdot D-\gamma_4\mu_f+m_f) \nonumber\\
&=& {\rm Det}\Big[(\tau_c^2C\gamma_5)^{-1}(\gamma\cdot D-\gamma_4\mu_f+m_f)(\tau_c^2C\gamma_5)\Big] \nonumber\\
&=& {\rm Det}(\gamma\cdot D-\gamma_4\mu_f+m_f)^\dagger\ ,
\end{eqnarray}
owing to the pseudoreality of $SU(2)$ group, $\tau_c^2\tau_c^a\tau_c^2 = -(\tau_c^a)^T$, etc., and thus Eq.~(\ref{DetProd}) turns out to be always positive definite. Therefore, the sign problem is harmless to our concerning $N_f=2+2$ QC$_2$D theory. The same conclusion is obtained in the presence of diquark sources straightforwardly by adopting the Nambu-Gorkov representation~\cite{Fukushima:2008su}.

Next, we examine chiral-symmetry properties of the Lagrangian~(\ref{QC2DLagrangian}). Introducing an eight-component spinor $\Psi = (\psi_R,\tilde{\psi}_L)$ with right-handed quarks $\psi_{R}$ and conjugate left-handed quarks $\tilde{\psi}_L = \tau_c^2\sigma^2\psi_L^*$ ($\psi_{R(L)} = \frac{1\pm \gamma_5}{2}\psi$ and $\sigma^2$ is a Pauli matrix in the Dirac space), in the Weyl representation Eq.~(\ref{QC2DLagrangian}) is rewritten into
\begin{eqnarray}
{\cal L}_{\rm QC_2D}^{N_f=2+2} &=& -\Psi^\dagger  i \sigma\cdot {\cal D}\Psi \nonumber\\
&& + \frac{1}{2} \Psi^T\sigma^2\tau_c^2{\bm M}^T\Psi + {\rm H.c.} \ ,  \label{QC2DLagrangian2}
\end{eqnarray}
with $\sigma_\mu=({\bm \sigma}, i)$. In this equation chemical potentials are incorporated via an extended covariant derivative
\begin{eqnarray}
{\cal D}_\mu = D_\mu-{\bm V}_\mu\ , \label{CovariantV}
\end{eqnarray}
where
\begin{eqnarray}
{\bm V}_\mu = (\mu_q {\bm J}_q + \mu_Q{\bm J}_Q)\delta_{\mu4} \label{VDec}
\end{eqnarray}
with
\begin{eqnarray}
{\bm J}_q = \left(
\begin{array}{cc}
\hat{I}_q & 0 \\
0 & -\hat{I}_q \\
\end{array}
\right)_{8\times8}\ , \ \ {\bm J}_Q = \left(
\begin{array}{cc}
\hat{I}_Q & 0 \\
0 & -\hat{I}_Q \\
\end{array}
\right)_{8\times8}\ .
\end{eqnarray}
In other words, the chemical potential ${\bm V}_\mu$ enters as a``$U(1)_q \times U(1)_Q$ gauge field'' associated with the light and heavy-quark numbers, respectively, whose transformation law is 
\begin{eqnarray}
{\bm V}_\mu \to U_{Qq}{\bm V}_\mu U_{Qq}^\dagger-i\partial_\mu U_{Qq}U_{Qq}^\dagger\ , \label{VTransform}
\end{eqnarray}
with $U_{Qq} = {\rm exp}(-i\theta_q{\bm J}_q - i\theta_Q{\bm J}_Q)$. Meanwhile, the quark mass matrix ${\bm M}$ reads
\begin{eqnarray}
{\bm M} = m_q{\bm E}_q + m_Q{\bm E}_Q\ , \label{MSymplectic}
\end{eqnarray}
with
\begin{eqnarray}
{\bm E}_q = \left(
\begin{array}{cc}
0 & \hat{I}_q \\
-\hat{I}_q & 0 \\
\end{array}
\right)_{8\times8}\ , \ \ {\bm E}_Q = \left(
\begin{array}{cc}
0 & \hat{I}_Q \\
-\hat{I}_Q & 0 \\
\end{array}
\right)_{8\times8}\ .
\end{eqnarray}

In the absence of ${\bm V}_\mu$ and ${\bm M}$ the Lagrangian~(\ref{QC2DLagrangian2}) obviously possesses an $SU(8)$ symmetry with respect to the extended flavor space. This symmetry is the so-called Pauli-G\"{u}rsey $SU(8)$ symmetry which is understood as an extension of $SU(4)_L\times SU(4)_R$ chiral symmetry~\cite{Pauli:1957voo,Gursey:1958fzy}. In this basis the chiral condensate $\langle\bar{\psi}\psi\rangle$ is written as $\langle\bar{\psi}\psi\rangle =(1/2)\langle\Psi^T\sigma^2\tau_c^2{\bm E}\Psi + {\rm H.c.}\rangle$ where ${\bm E}$ is the $8\times8$ symplectic matrix 
\begin{eqnarray}
{\bm E} = {\bm E}_q + {\bm E}_Q = \left(
\begin{array}{cc}
0 & {\bm 1} \\
-{\bm 1} & 0 \\
\end{array}
\right)_{8\times8}\ .
\end{eqnarray}
 As a result the breaking pattern of the Pauli-G\"{u}rsey symmetry turns out to be $SU(8) \to Sp(8)$, and accordingly $27$ NG bosons emerge in the chiral limit. Therefore, as long as quark masses $m_q$ and $m_Q$ are not sufficiently large, effective models describing low-lying hadrons in $N_f=2+2$ QC$_2$D should be constructed upon the approximate symmetry breaking of $SU(8)\to Sp(8)$. The chemical potentials in effective models are turned on by simply covariantizing derivatives appropriately as in Eq.~(\ref{CovariantV})~\cite{Suenaga:2025sln}.

\subsection{Construction of the LSM}
\label{sec:LSM}

In this subsection we introduce our LSM in QC$_2$D with $N_f=2+2$ based on symmetry properties explained in Sec.~\ref{sec:QC2DLagrangian}. In what follows, we will adopt the Minkowskian notation.

Since the linear representation of the Pauli-G\"{u}rsey $SU(8)$ symmetry plays a central role in the LSM framework, we introduce the following quark bilinear~\cite{Suenaga:2019jjv,Suenaga:2025sln}:
\begin{eqnarray}
\Sigma_{ij} \sim \Psi_j^T\sigma^2\tau^2_c\Psi_i\ , \label{SigmaBilinear}
\end{eqnarray}
with $i,j=1$ - $8$, the $SU(8)$ properties of which are transparent. $\Sigma$ is an anti-symmetric $8\times8$ matrix: $\Sigma^T=-\Sigma$ which has $28$ components, so that it can be parametrized by generators of $G/H=SU(8)/Sp(8)$, $X^{a=1-27}$, and a trivial matrix $X^{a=0}$, listed in Appendix~\ref{sec:Generators}. Making use of the property~(\ref{XProperty}) and Grassmann nature of $\Psi$, with the loss of generality $\Sigma$ takes the form of
\begin{eqnarray}
\Sigma = ({\cal S}^a-i{\cal P}^a)X^a {\bm E}\ . \label{SigmaDef}
\end{eqnarray}
The ${\cal P}^a$ and ${\cal S}^a$ correspond to spin-$0$ hadron fields, the quantum numbers of which are summarized in Table~\ref{tab:HadronP} and Table~\ref{tab:HadronS}. In these tables, $J^P$, $N_q$, $N_Q$, $I_q$ and $I_Q$ stand for spin and parity, light quark number, heavy quark number, light isospin and heavy isospin, respectively.
 

\begin{table*}[t]
\begin{center}
  \begin{tabular}{c||c|ccccc}  \hline\hline
Hadron & Field parametrization & $J^P$ & $N_q$ & $N_Q$ & $I_q$ & $I_Q$ \\ \hline
$\eta_{qq}$ & $ \frac{1}{\sqrt{2}}{\cal P}^0 + \frac{1}{\sqrt{3}}{\cal P}^8 + \frac{1}{\sqrt{6}}{\cal P}^{15}$ &  $0^-$ & $0$ & $0$ & 0 & 0\\ 
$\eta_{QQ}$ & $-\frac{1}{\sqrt{3}}{\cal P}^8+\sqrt{\frac{2}{3}}{\cal P}^{15}$ &  $0^-$ & $0$ & $0$ & 0 & 0\\ 
$\pi_{0,qq}$ & ${\cal P}^{3}$,  $\frac{1}{\sqrt{2}}({\cal P}^1 \pm i{\cal P}^2)$ & $0^-$ & $0$ & $0$ & $1$  & $0$\\
$\pi_{0,QQ}$ & $ \frac{1}{\sqrt{2}}{\cal P}^0-\frac{1}{\sqrt{3}}{\cal P}^8-\frac{1}{\sqrt{6}}{\cal P}^{15}$, $ \frac{1}{\sqrt{2}}({\cal P}^{13}\pm i{\cal P}^{14})$ & $0^-$ & $0$ & $0$ & $0$  & $1$\\
$K_{Qq}$ & $ \frac{1}{\sqrt{2}}({\cal P}^4- i{\cal P}^5)$, $ \frac{1}{\sqrt{2}}({\cal P}^6- i{\cal P}^7) $, $ \frac{1}{\sqrt{2}}({\cal P}^9- i{\cal P}^{10})$, $ \frac{1}{\sqrt{2}}({\cal P}^{11}- i{\cal P}^{12}) $ & $0^-$ & $+1$ & $-1$ & $\frac{1}{2}$ & $\frac{1}{2}$ \\
$\bar{K}_{Qq}$ & $ \frac{1}{\sqrt{2}}({\cal P}^4+ i{\cal P}^5)$, $ \frac{1}{\sqrt{2}}({\cal P}^6 + i{\cal P}^7) $, $ \frac{1}{\sqrt{2}}({\cal P}^9 + i{\cal P}^{10})$, $ \frac{1}{\sqrt{2}}({\cal P}^{11} + i{\cal P}^{12}) $ & $0^-$ & $-1$ & $+1$ & $\frac{1}{2}$ & $\frac{1}{2}$ \\
$B_{qq}$ & $\frac{1}{\sqrt{2}}({\cal P}^{17}-i{\cal P}^{16})$ & $0^+$ & $+2$ & $0$ & $0$ & $0$ \\
$\bar{B}_{qq}$ & $\frac{1}{\sqrt{2}}({\cal P}^{17}+i{\cal P}^{16})$ & $0^+$ & $-2$ & $0$ & $0$ & $0$ \\
$B_{Qq}$ & $\frac{1}{\sqrt{2}}({\cal P}^{19}-i{\cal P}^{18})$, $\frac{1}{\sqrt{2}}({\cal P}^{21}-i{\cal P}^{20})$, $\frac{1}{\sqrt{2}}({\cal P}^{23}-i{\cal P}^{22})$, $\frac{1}{\sqrt{2}}({\cal P}^{25}-i{\cal P}^{24})$ & $0^+$ & $+1$ & $+1$ & $0$ & $0$ \\
$\bar{B}_{Qq}$ & $\frac{1}{\sqrt{2}}({\cal P}^{19}+i{\cal P}^{18})$, $\frac{1}{\sqrt{2}}({\cal P}^{21}+i{\cal P}^{20})$, $\frac{1}{\sqrt{2}}({\cal P}^{23}+i{\cal P}^{22})$, $\frac{1}{\sqrt{2}}({\cal P}^{25}+i{\cal P}^{24})$ & $0^+$ & $-1$ & $-1$ & $0$ & $0$ \\
$B_{QQ}$ & $\frac{1}{\sqrt{2}}({\cal P}^{27}-i{\cal P}^{26})$ & $0^+$ & $0$ & $+2$ & $0$ & $0$ \\
$\bar{B}_{QQ}$ & $\frac{1}{\sqrt{2}}({\cal P}^{27}+i{\cal P}^{26})$ & $0^+$ & $0$ & $-2$ & $0$ & $0$ \\
\hline \hline
 \end{tabular}
\caption{Quantum numbers of the hadrons generated by ${\cal P}^a$.}
\label{tab:HadronP}
\end{center}
\end{table*}

\begin{table*}[t]
\begin{center}
  \begin{tabular}{c||c|ccccc}  \hline\hline
Hadron & Field parametrization & $J^P$ & $N_q$ & $N_Q$ & $I_q$ & $I_Q$ \\ \hline
$\sigma_{qq}$ & $ \frac{1}{\sqrt{2}}{\cal S}^0 + \frac{1}{\sqrt{3}}{\cal S}^8 + \frac{1}{\sqrt{6}}{\cal S}^{15}$ &  $0^+$ & $0$ & $0$ & 0 & 0\\ 
$\sigma_{QQ}$ & $-\frac{1}{\sqrt{3}}{\cal S}^8+\sqrt{\frac{2}{3}}{\cal S}^{15}$ &  $0^+$ & $0$ & $0$ & 0 & 0\\ 
$a_{0,qq}$ & ${\cal S}^{3}$,  $\frac{1}{\sqrt{2}}({\cal S}^1 \pm i{\cal S}^2)$ & $0^+$ & $0$ & $0$ & $1$  & $0$\\
$a_{0,QQ}$ & $ \frac{1}{\sqrt{2}}{\cal S}^0-\frac{1}{\sqrt{3}}{\cal S}^8-\frac{1}{\sqrt{6}}{\cal S}^{15}$, $ \frac{1}{\sqrt{2}}({\cal S}^{13}\pm i{\cal S}^{14})$ & $0^+$ & $0$ & $0$ & $0$  & $1$\\
$\kappa_{Qq}$ & $ \frac{1}{\sqrt{2}}({\cal S}^4- i{\cal S}^5)$, $ \frac{1}{\sqrt{2}}({\cal S}^6- i{\cal S}^7) $, $ \frac{1}{\sqrt{2}}({\cal S}^9- i{\cal S}^{10})$, $ \frac{1}{\sqrt{2}}({\cal S}^{11}- i{\cal S}^{12}) $ & $0^+$ & $+1$ & $-1$ & $\frac{1}{2}$ & $\frac{1}{2}$ \\
$\bar{\kappa}_{Qq}$ & $ \frac{1}{\sqrt{2}}({\cal S}^4+ i{\cal S}^5)$, $ \frac{1}{\sqrt{2}}({\cal S}^6 + i{\cal S}^7) $, $ \frac{1}{\sqrt{2}}({\cal S}^9 + i{\cal S}^{10})$, $ \frac{1}{\sqrt{2}}({\cal S}^{11} + i{\cal S}^{12}) $ & $0^+$ & $-1$ & $+1$ & $\frac{1}{2}$ & $\frac{1}{2}$ \\
$B'_{qq}$ & $\frac{1}{\sqrt{2}}({\cal S}^{17}-i{\cal S}^{16})$ & $0^-$ & $+2$ & $0$ & $0$ & $0$ \\
$\bar{B}'_{qq}$ & $\frac{1}{\sqrt{2}}({\cal S}^{17}+i{\cal S}^{16})$ & $0^-$ & $-2$ & $0$ & $0$ & $0$ \\
$B'_{Qq}$ & $\frac{1}{\sqrt{2}}({\cal S}^{19}-i{\cal S}^{18})$, $\frac{1}{\sqrt{2}}({\cal S}^{21}-i{\cal S}^{20})$, $\frac{1}{\sqrt{2}}({\cal S}^{23}-i{\cal S}^{22})$, $\frac{1}{\sqrt{2}}({\cal S}^{25}-i{\cal S}^{24})$ & $0^-$ & $+1$ & $+1$ & $0$ & $0$ \\
$\bar{B}'_{Qq}$ & $\frac{1}{\sqrt{2}}({\cal S}^{19}+i{\cal S}^{18})$, $\frac{1}{\sqrt{2}}({\cal S}^{21}+i{\cal S}^{20})$, $\frac{1}{\sqrt{2}}({\cal S}^{23}+i{\cal S}^{22})$, $\frac{1}{\sqrt{2}}({\cal S}^{25}+i{\cal S}^{24})$ & $0^-$ & $-1$ & $-1$ & $0$ & $0$ \\
$B'_{QQ}$ & $\frac{1}{\sqrt{2}}({\cal S}^{27}-i{\cal S}^{26})$ & $0^-$ & $0$ & $+2$ & $0$ & $0$ \\
$\bar{B}'_{QQ}$ & $\frac{1}{\sqrt{2}}({\cal S}^{27}+i{\cal S}^{26})$ & $0^-$ & $0$ & $-2$ & $0$ & $0$ \\
\hline \hline
 \end{tabular}
\caption{Quantum numbers of the hadrons generated by ${\cal S}^a$.}
\label{tab:HadronS}
\end{center}
\end{table*}

The linear representation~(\ref{SigmaBilinear}) implies that $\Sigma$ is transformed as 
\begin{eqnarray}
\Sigma \to U \Sigma U^T \ \ \ {\rm with}\ \ \ U\in SU(8)\ , \label{SigmaTrans}
\end{eqnarray}
under the Pauli-G\"{u}rsey $SU(8)$ transformation. Hence an $SU(8)$ symmetric Lagrangian up to ${\cal O}(\Sigma^4)$ reads
\begin{eqnarray}
{\cal L}_{\rm LSM} &=& {\rm tr}[D_\mu\Sigma^\dagger D^\mu\Sigma] + \bar{c}{\rm tr}[\zeta^\dagger\Sigma + \zeta \Sigma^\dagger] \nonumber\\
&& -m_0^2{\rm tr}[\Sigma^\dagger\Sigma]- \lambda_1\left({\rm tr}[\Sigma^\dagger\Sigma]\right)^2-\lambda_2{\rm tr}\left[\left(\Sigma^\dagger\Sigma\right)^2\right] \nonumber\\
&& + c\, \epsilon_{ijklmnop}\left( \Sigma_{ij}\Sigma_{kl}\Sigma_{mn} \Sigma_{op}+{\rm H.c.} \right)\ , \label{LSM8}
\end{eqnarray}
where the covariant derivative $D_\mu\Sigma = \partial_\mu-i({\bm V}_\mu \Sigma + {\bm V}_\mu^T\Sigma)$ is introduced for which the chemical potentials in Eq.~(\ref{VDec}) are incorporated appropriately from the $U(1)_q\times U(1)_Q$ transformation property~(\ref{VTransform}). The spurious field $\zeta$ in Eq.~(\ref{LSM8}) transforming as $\zeta \to U\zeta U^T$ is introduced so as to describe the explicit breaking of $SU(8) \to Sp(8)$ in a systematic way. Thus, this $\zeta$ is replaced by its vacuum expectation value
\begin{eqnarray}
 \langle \zeta\rangle = \frac{1}{2}{\bm M}\ ,
 \end{eqnarray}
 with Eq.~(\ref{MSymplectic}) in the end. The last term of Eq.~(\ref{LSM8}) is invariant under Eq.~(\ref{SigmaTrans}) since all the indices are contracted anti-symmetrically to yield $SU(8)$ singlet, but only breaks a $U(1)$ symmetry generated by $\Sigma \to {\rm exp}(-2i\theta_A)\Sigma$ [see Eq.~(\ref{SigmaBilinear})]. This is nothing but the $U(1)$ axial transformation, and hence the $c$ term is responsible for the $U(1)$ axial anomaly effects.

Since the chiral condensate is a $J^P=0^+$ operator carrying no quark numbers or isospins, the spontaneous breakdown of $SU(8)$ is triggered by emergence of mean fields of $\sigma_{qq}$ and $\sigma_{QQ}$: 
\begin{eqnarray}
\phi_{q} \equiv\langle\sigma_{qq}\rangle\ ,  \ \ \phi_{Q} \equiv\langle\sigma_{QQ}\rangle\ , \label{PhiMean}
\end{eqnarray}
within the present LSM framework. Besides, in principle all the $J^P=0^+$ diquarks corresponding to $B_{qq}$, $B_{Qq}$ and $B_{QQ}$ are capable of being condensed when the chemical potentials are adequately large. 
In the following analysis,  we will present demonstrations with which only the mean field of $B_{qq}$ is generated.\footnote{As long as we stick to small $\mu_q$ regime this assumption is correct. Possibility of emergence of the heavy diquark condensates $\langle B_{QQ}\rangle$ from viewpoints of the $U(1)_A$ anomaly effects will be discussed at the end of Sec.~\ref{sec:HadronMass}.} When choosing a phase such that the diquark condensate is pure real, the mean field is given by
\begin{eqnarray}
\Delta_{q} \equiv\langle {\cal P}^{17}\rangle\ , \label{DeltaMean}
\end{eqnarray}
from Table~\ref{tab:HadronP}.

\begin{figure}[t]
\centering
\hspace*{-0.5cm} 
\includegraphics*[scale=0.65]{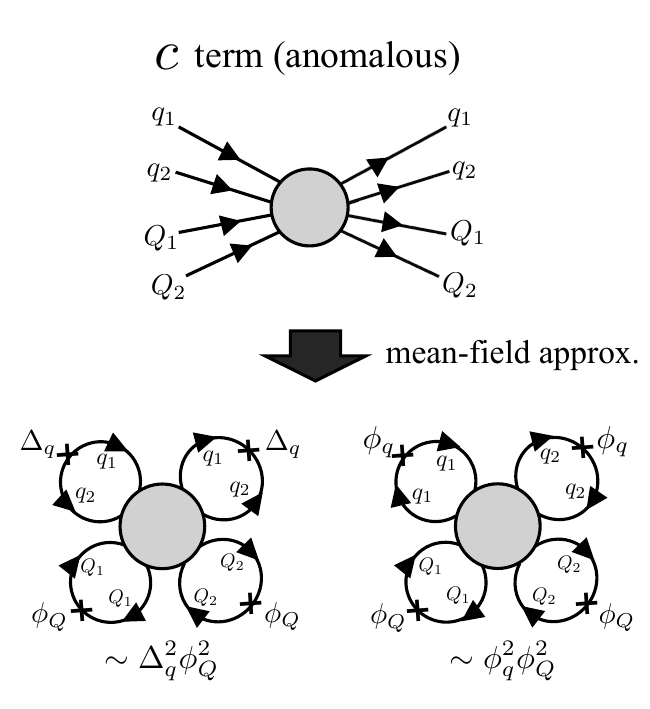}
\caption{Top: the eight-point interaction generated by the anomalous $c$ term. Bottom: two types of the anomalous contributions to the potential $V_{\rm LSM}$.}
\label{fig:Anomaly}
\end{figure}

Substituting Eq.~(\ref{SigmaDef}) into the Lagrangian~(\ref{LSM8}) with the field parametrization in Table.~\ref{tab:HadronP} and Table.~\ref{tab:HadronS}, and only leaving the mean fields~(\ref{PhiMean}) and~(\ref{DeltaMean}), a potential $V_{\rm LSM}$ at mean-field level is obtained to be
\begin{eqnarray}
&& V_{\rm LSM} = -2\mu_{q}\phi_{q}\Delta_{q}^2-\sqrt{2}\bar{c}\left(m_q\phi_{q} + m_Q\phi_{Q}\right) \nonumber\\
&&\ \ \ \  + \frac{m_0^2}{2}\left(\Delta_{q}^2+\phi_{q}^2+\phi_{Q}^2\right)  +\frac{\lambda_1}{4}\left(\Delta_{q}^2+\phi_{q}^2+\phi_{Q}^2\right)^2 \nonumber\\
&& \ \ \ \ +\frac{\lambda_2}{16}\Big[\left(\Delta_{q}^2+\phi_{q}^2\right)^2+\phi_{Q}^4\Big]-12c\left(\phi_{q}^2 + \Delta_{q}^2\right)\phi_{Q}^2\ .\nonumber\\ \label{VLSM}
\end{eqnarray}
The actual ground-state configurations are determined by solving the respective stationary conditions: $\partial V_{\rm LSM}/\partial\phi_{q}=0$, $\partial V_{\rm LSM}/\partial\phi_{Q}=0$, $\partial V_{\rm LSM}/\partial\Delta_{q}=0$. For later convenience we present the equations explicitly here
\begin{eqnarray}
&& m_0^2 + \lambda_1\left(\Delta_{q}^2+\phi_{q}^2+\phi_{Q}^2\right) +\frac{\lambda_2}{4}\left(\Delta_{q}^2+\phi_{q}^2\right) \nonumber\\
&& -24c\phi_{Q}^2 -\frac{\sqrt{2}\bar{c}m_q}{\phi_{q}}=0 \ , \label{PhiqqGEq}
\end{eqnarray}
\begin{eqnarray}
&& m_0^2 +\lambda_1\phi_{Q}^2+ \frac{\lambda_2}{4}\phi_{Q}^2 -\frac{\sqrt{2}\bar{c}{m}_Q}{\phi_{Q}} \nonumber\\
&& = (24c-\lambda_1)(\Delta_{q}^2+\phi_{q}^2)  \ , \label{PhiQQGEq}
\end{eqnarray}
\begin{eqnarray}
&&\Bigg[-4\mu_{q}^2 +m_0^2+ \lambda_1\left(\Delta_{q}^2+\phi_{q}^2+\phi_{Q}^2\right) \nonumber\\
&& + \frac{\lambda_2}{4}\left(\Delta_{q}^2+\phi_{q}^2\right) - 24c\phi_{Q}^2\Bigg] \Delta_{q} = 0\ , \label{DeltaqqGEq}
\end{eqnarray}
in which $\phi_q\neq 0$ and $\phi_Q\neq 0$ are naturally assumed.

Hadron mass formulas in cold matter are achieved by expanding the Lagrangian up to ${\cal O}(\Sigma^2)$ upon the mean fields. In order to avoid lengthy equations we collectively exhibit them in Appendix~\ref{sec:MassFormulas}, instead, here we pick up important points of the mass formulas. With finite chemical potentials, time derivatives of some hadrons mix with $\mu_q$ or $\mu_Q$ accordingly to their quark numbers. In such cases we need to find pole positions of the propagator matrix (in momentum space) at the rest so as to correctly evaluate their masses. When the chemical potential $\mu_q$ exceeds a critical value (which will be turned out to be one-half of the vacuum pion mass), diquark condensates $\Delta_{q}$ indeed emerge which leads to the baryon superfluid phase~\cite{Kogut:1999iv,Kogut:2000ek}. In this phase, mesons and (anti)diquark baryons would mix owing to the light-quark number ($N_q$) violation, and thus it is inevitable to seek for pole positions of the further enlarged propagator matrix. We note that the diquark condensates keep all the quantum numbers conserved other than $N_q$. Therefore the mixings occur among
\begin{eqnarray}
&& (\eta_{qq},\eta_{QQ}, B'_{qq},\bar{B}'_{qq})\ , \ \ (\sigma_{qq},\sigma_{QQ}, B_{qq},\bar{B}_{qq})\ , \nonumber\\
&& (K_{Qq},\bar{K}_{Qq}, B'_{Qq},\bar{B}'_{Qq})\ ,\ \  (\kappa_{Qq},\bar{\kappa}_{Qq}, B_{Qq},\bar{B}_{Qq})\ . \nonumber\\ \label{MixingStructure}
\end{eqnarray}
We also note that even in the hadronic phase, $\sigma_{qq}$ and $\sigma_{QQ}$ or $\eta_{qq}$ and $\eta_{QQ}$ can mix via $c$ and $\lambda_1$ couplings.

\subsection{Role of the anomalous $c$ term}
\label{sec:CTerm}

Before moving on to numerical analyses, here we explain properties of the anomalous $c$ term which will play crucial roles in determining the fate of mean fields in dense medium.

\begin{figure}[t]
\centering
\hspace*{-0.2cm} 
\includegraphics*[scale=0.6]{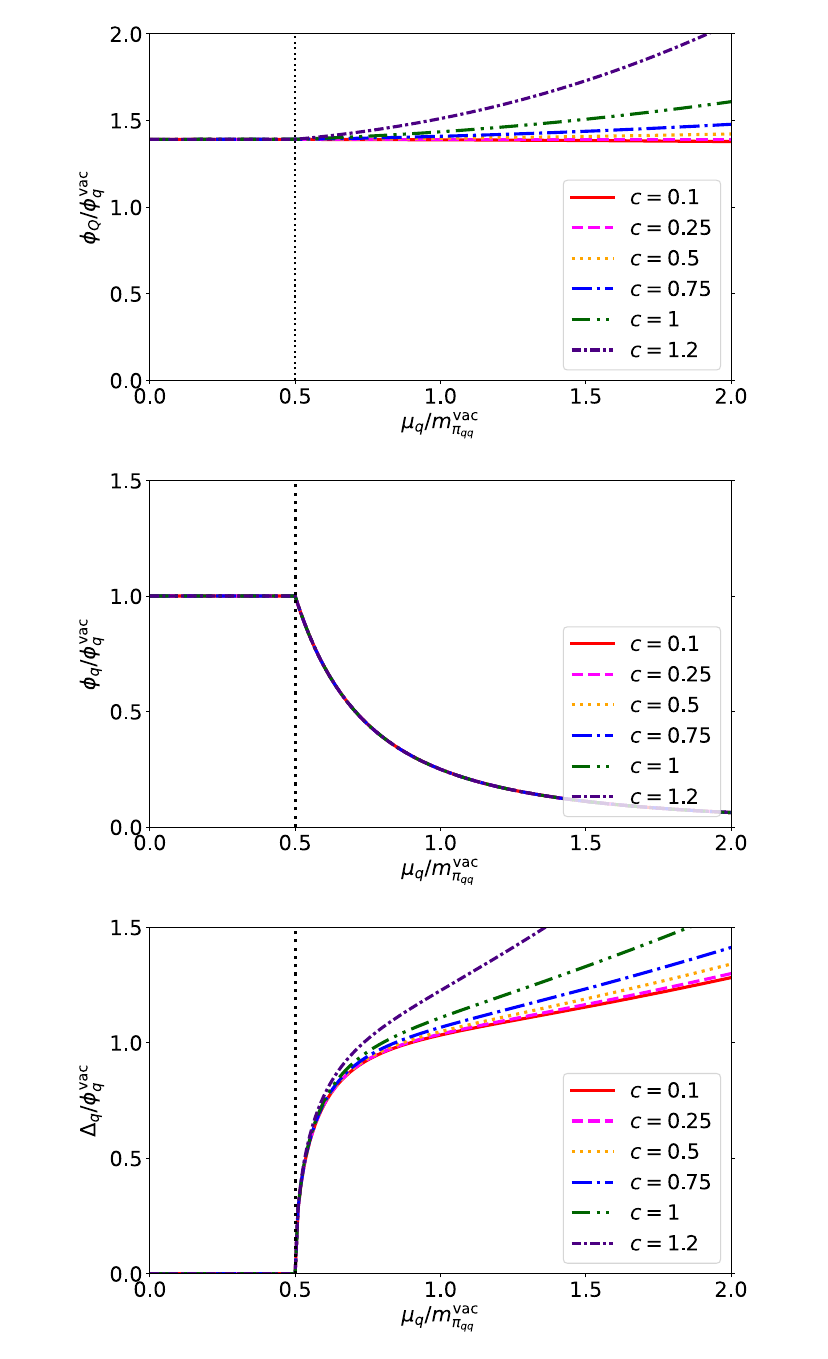}
\caption{$\mu_q$ dependences of $\phi_Q$ with several values of $c$ (top). We also present those of $\phi_q$ (middle) and $\Delta_q$ (bottom) for completeness. The vertical dotted line indicated $\mu_q^{\rm cr} = m_{\pi_{qq}}^{\rm vac}/2$.}
\label{fig:ThreeMFs}
\end{figure}

\begin{figure}[t]
\centering
\hspace*{-0.5cm} 
\includegraphics*[scale=0.3]{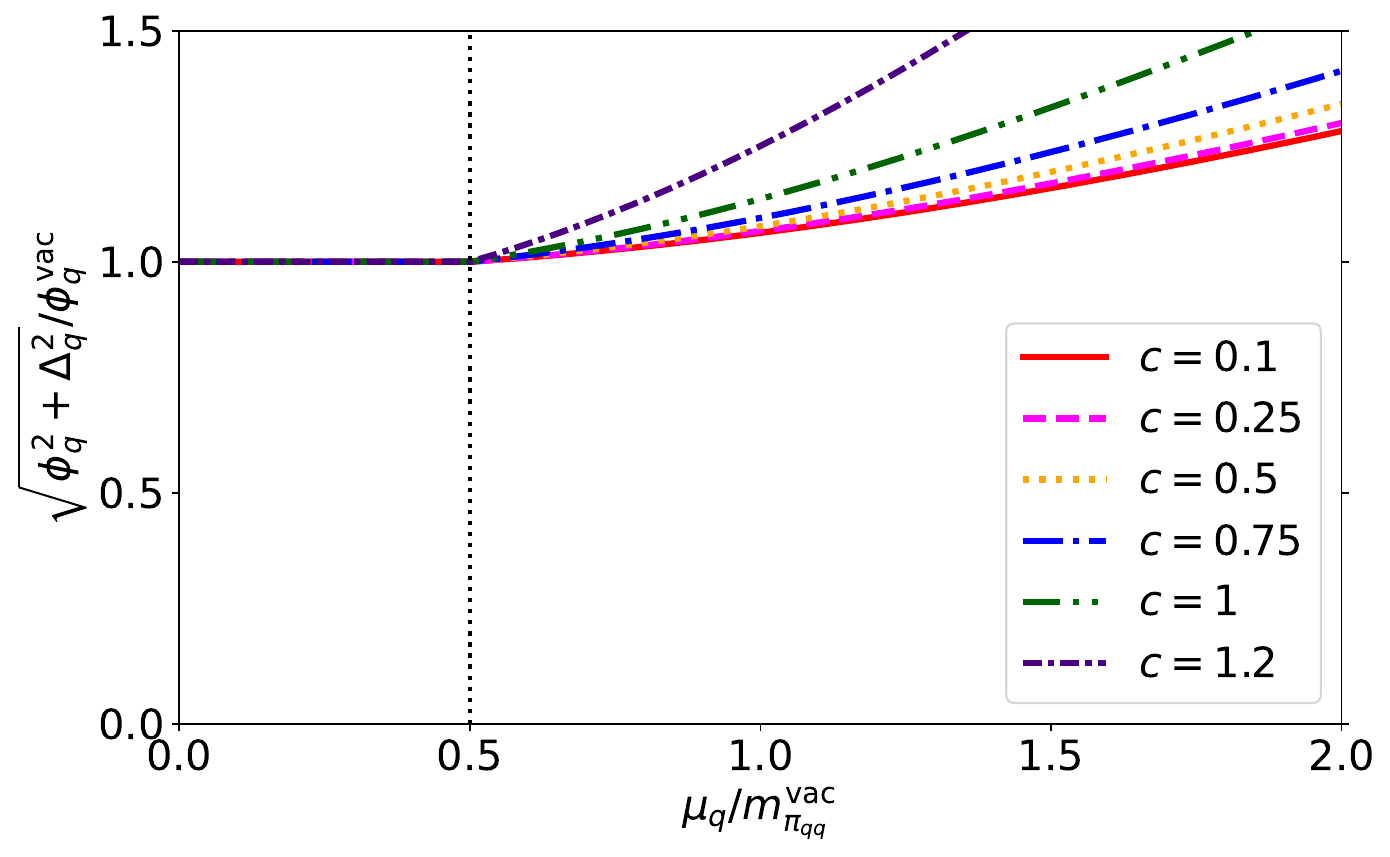}
\caption{$\mu_q$ dependences of $\sqrt{\phi_q^2+\Delta_q^2}$ with several values of $c$.}
\label{fig:DeltaPhiq}
\end{figure}

In the $c$ term all indices of $\Sigma$ are contracted anti-symmetrically with the $\epsilon$ tensor, which implies that all the indices must carry different quark flavors~\cite{Kobayashi:1970ji,Kobayashi:1971qz,tHooft:1976snw,tHooft:1976rip}, from the bilinear representation~(\ref{SigmaBilinear}). At quark-line level, schematically the anomalous interactions are represented in the top eight-point interaction of Fig.~\ref{fig:Anomaly}. Hence, within the mean-field approximation this term generates two-types of the contributions proportional to $\Delta_{q}^2\phi_{Q}^2$ and $\phi_{q}^2\phi_{Q}^2$, as depicted in the bottom two diagrams in Fig.~\ref{fig:Anomaly}. They indeed exist in the last term of the potential~(\ref{VLSM}). In particular, the direct coupling between the light diquark condensates and heavy chiral condensates, $\Delta_q^2$ and $\phi_{Q}^2$, is regarded as a characteristic feature of cold and dense QC$_2$D with $N_f=2+2$. 

Anomalous contributions to the hadron masses are read off by cutting two loops in the bottom diagrams of Fig.~\ref{fig:Anomaly}. Thus the flavor-mixing structures also apply to the masses, which are indeed reflected by the mass formulas summarized in Appendix.~\ref{sec:MassFormulas}.

Finally, we note that similar flavor-mixing couplings are driven by the $\lambda_1$ term,  since the quark lines are separated due to its double-trace property. Both the $\lambda_1$ and $c$ couplings are expected to be of ${\cal O}(N_c^{-2})$ from the 't Hooft large-$N_c$ counting, which is suppressed compared to $\lambda_2\sim {\cal O}(N_c^{-1})$~\cite{Witten:1979kh}. 


\section{Numerical results}
\label{sec:Numerical}

In this section we present numerical results by means of the present $N_f=2+2$ LSM particularly focusing on the $U(1)$ axial anomaly effects. First, in Sec.~\ref{sec:MeanField} we explore the fate of chiral and diquark condensates at finite quark chemical potential within a mean-field approximation. Next, in Sec.~\ref{sec:HadronMass} numerical results on the hadron mass spectrum in cold and dense medium are shown. Then, in Sec.~\ref{sec:Topological} we investigate the topological susceptibility from the viewpoints of chiral-symmetry restoration.

\subsection{Mean fields at finite $\mu_q$}
\label{sec:MeanField}

Here we show phase structures of cold and dense QC$_2$D with $N_f=2+2$ by investigating chemical-potential dependences of the mean fields.

Before depicting numerical results we explain general properties of the mean fields. From the light-pion mass formula~(\ref{PiqqMass}), Eq.~(\ref{DeltaqqGEq}) can be rewritten to
\begin{eqnarray}
(-4\mu_q^2+m_{\pi_{qq}}^2) \Delta_{q}= 0\ . \label{DeltaCond}
\end{eqnarray}
As long as $\Delta_{q}=0$ the stationary conditions for $\phi_{q}$ and $\phi_{Q}$, Eqs.~(\ref{PhiqqGEq}) and~(\ref{PhiQQGEq}), do not change from the vacuum ones, and so does the light-pion mass $m_{\pi_{qq}}$. Hence, in order to connect the $\pi_{qq}$ mass continuously at any $\mu_q$, one can analytically confirm that the ground-state configuration is separated by a critical chemical potential $\mu_q^{\rm cr} \equiv m^{\rm vac}_{\pi_{qq}}/2$\footnote{The superscript ``vac'' is attached to stress that the quantity is defined in the vacuum ($\mu_q=0$).} as
\begin{eqnarray}
&& \Delta_{q} = 0\ \ \ {\rm for}\ \ \mu_{q}< \mu_{q}^{\rm cr}\ \ \  ({\rm Hadronic\ phase})\nonumber\\
&&  \Delta_{q} \neq 0\ \ \ {\rm for}\ \ \mu_{q}\geq \mu_{q}^{\rm cr}  \ \  ({\rm Baryon\ superfluid\ phase})\ , \nonumber\\
\end{eqnarray}
from Eq.~(\ref{DeltaCond}), which is identical to the scenario with $N_f=2$~\cite{Kogut:2000ek,Ratti:2004ra,Suenaga:2022uqn}. Following the previous studies we call the hadronic phase for $\mu_{q}< \mu_{q}^{\rm cr}$ and the baryon superfluid phase for $\mu_{q}\geq \mu_{q}^{\rm cr}$.


\begin{figure*}[t]
\centering
\hspace*{-0.5cm} 
\includegraphics*[scale=0.65]{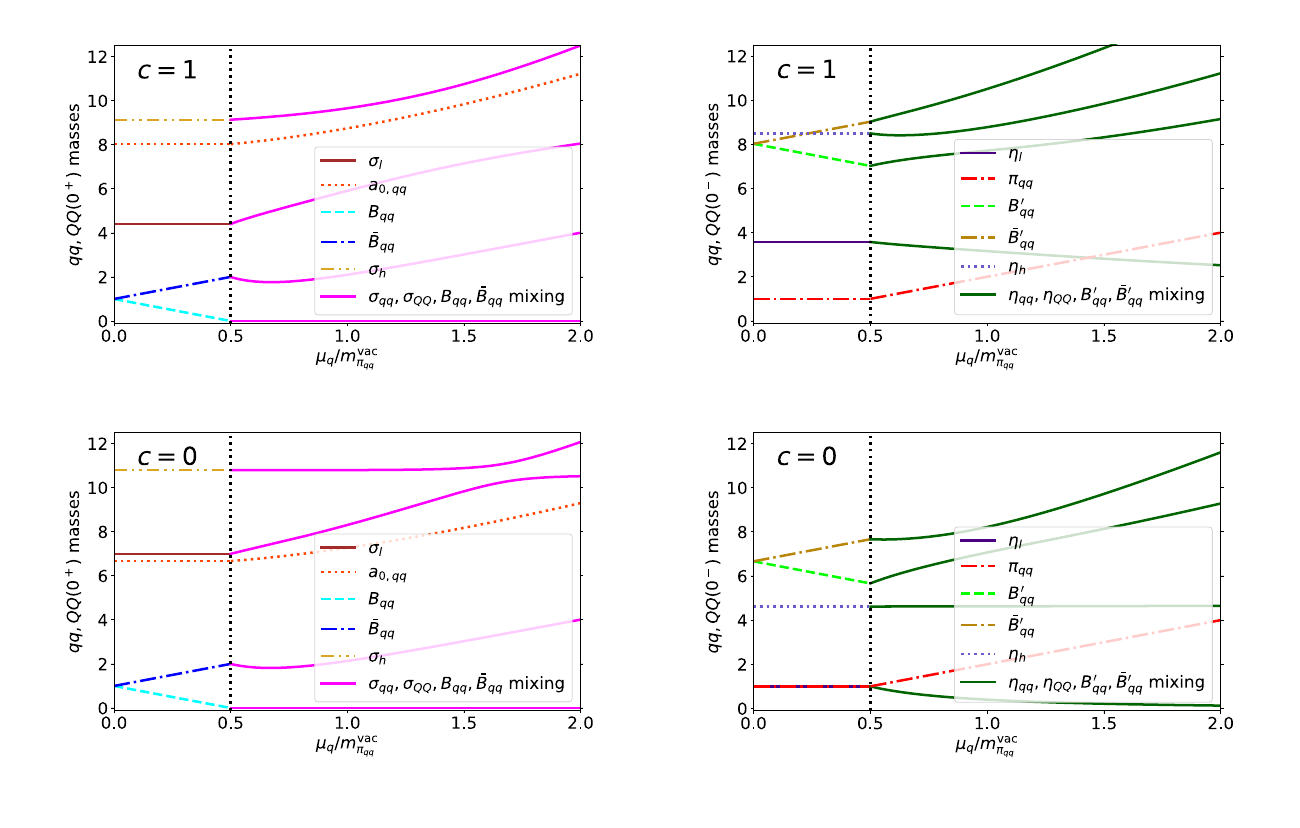}
\caption{Mass spectra of light $0^+$ (left) and $0^-$ (right) hadrons at finite $\mu_q$ ($\mu_q=\mu_Q$), with the parameter set I with $c=1$ (top) and set II with $c=0$ (bottom) in Table~\ref{tab:Parameters}. The masses are normalized by $m_{\pi_{qq}}^{\rm vac}$. The masses of $\sigma_{h}$ and $\eta_{h}$ are also exhibited in these panels owing to the mixings.}
\label{fig:MassLLHH}
\end{figure*}
\begin{figure*}[t]
\centering
\hspace*{-0.5cm} 
\includegraphics*[scale=0.65]{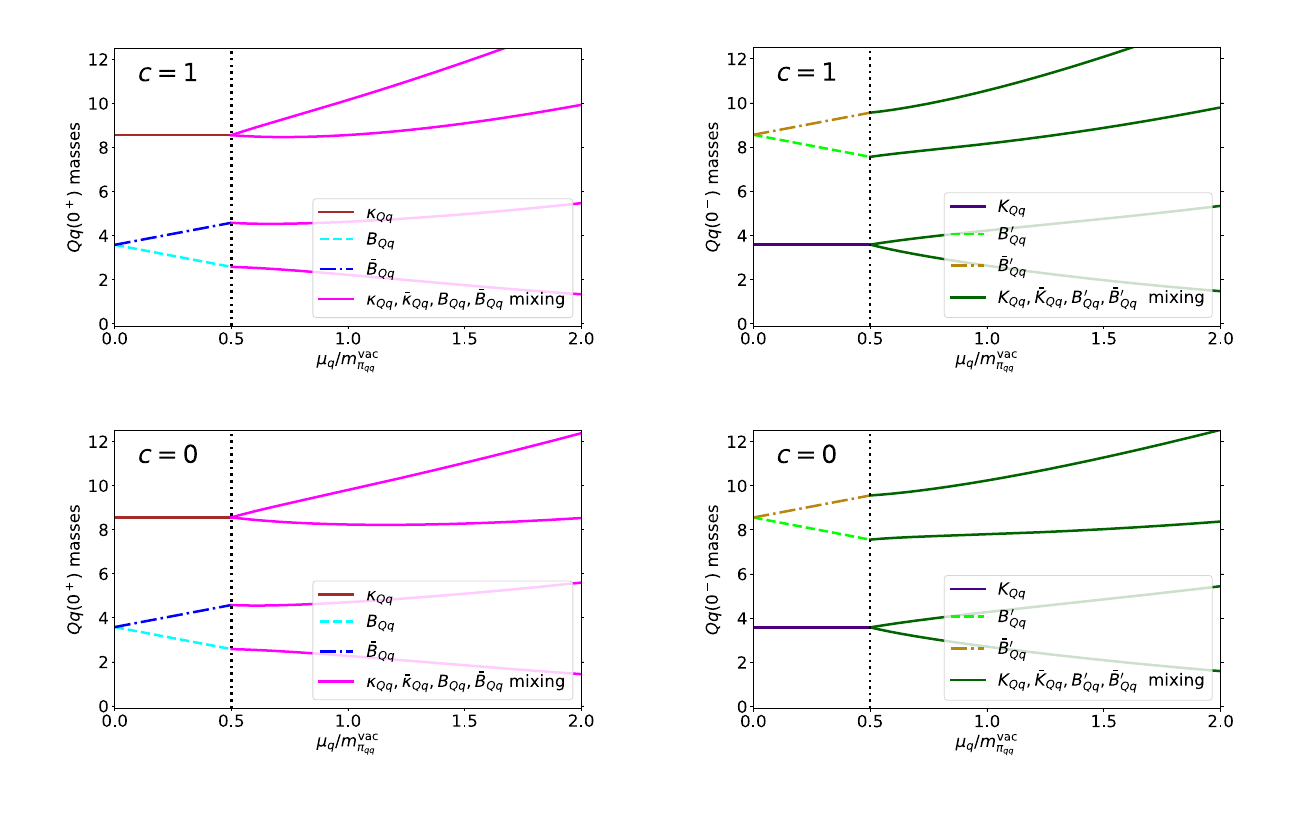}
\caption{Mass spectra of heavy-light $0^+$ (left) and $0^-$ (right) hadrons at finite $\mu_q$ ($\mu_q=\mu_Q$). The same parameters as in Fig.~\ref{fig:MassLLHH} are adopted.}
\label{fig:MassHL}
\end{figure*}
\begin{figure*}[t]
\centering
\hspace*{-0.5cm} 
\includegraphics*[scale=0.65]{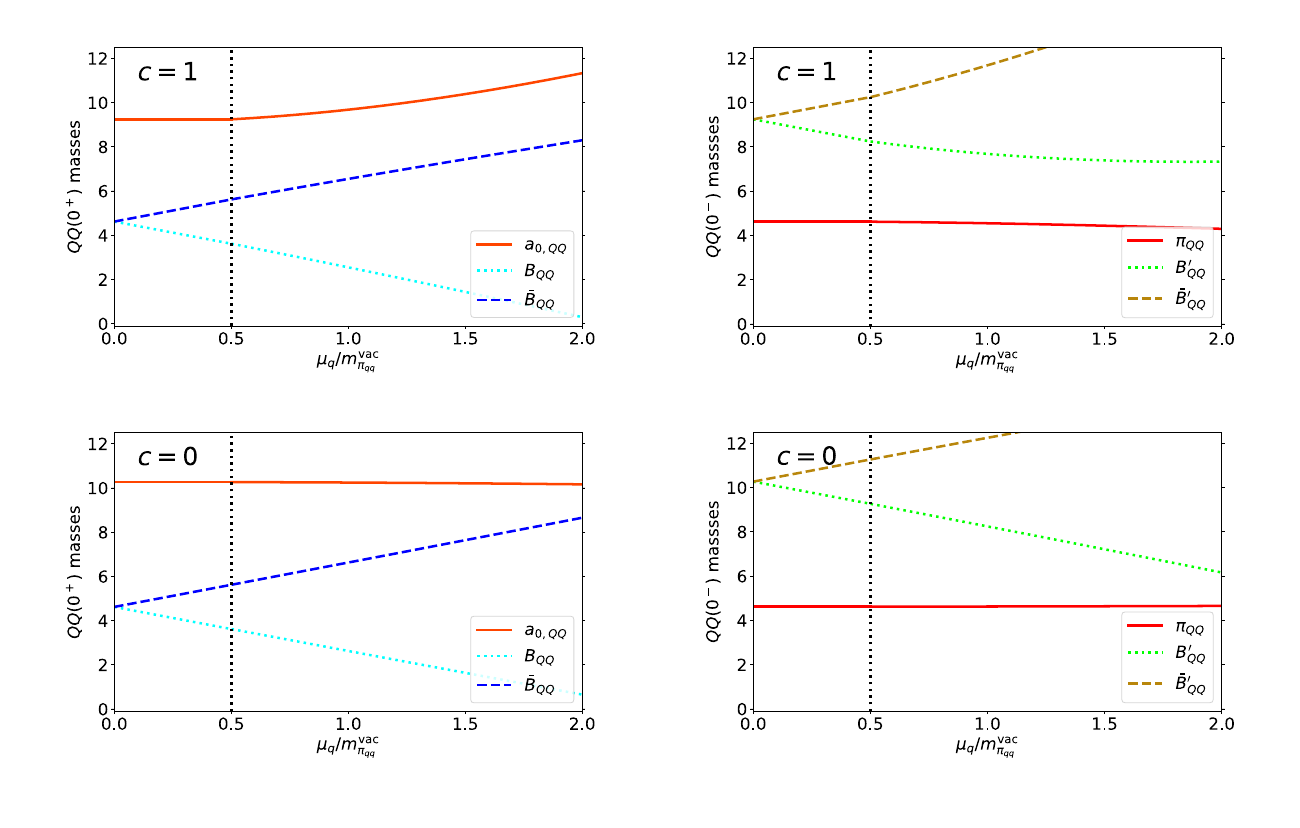}
\caption{Mass spectra of the remaining heavy $0^+$ (left) and $0^-$ (right) hadrons at finite $\mu_q$ ($\mu_q=\mu_Q$). The same parameters as in Fig.~\ref{fig:MassLLHH} are adopted.}
\label{fig:MassHH}
\end{figure*}

\begin{figure}[t]
\centering
\hspace*{-0.5cm} 
\includegraphics*[scale=0.3]{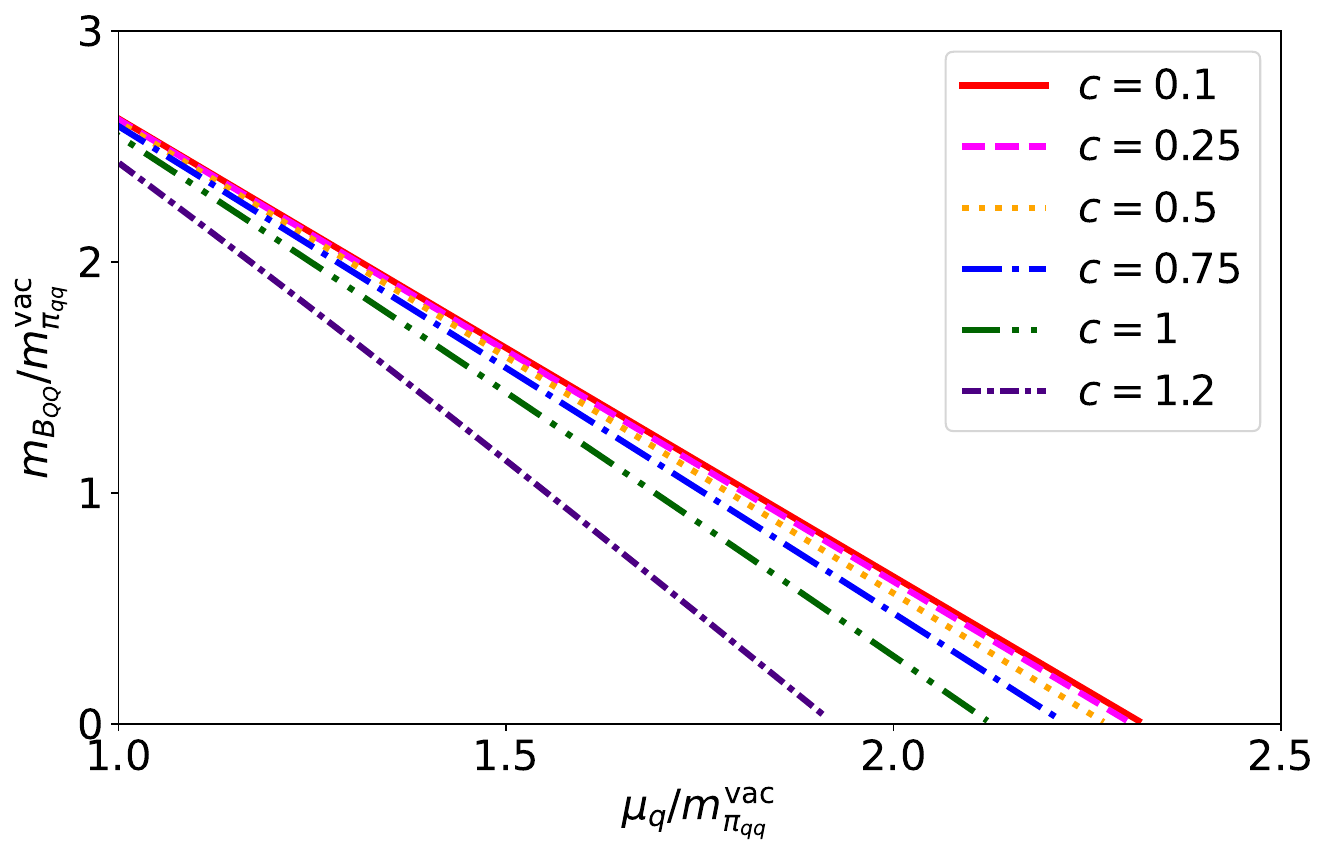}
\caption{$\mu_q$ dependences of the $B_{QQ}$ mass ($m_{B_{QQ}}$) focusing on the zero crossing.}
\label{fig:FateOfBQQ}
\end{figure}

In the hadronic phase, the chiral condensates are identical to the vacuum ones: $\phi_{q}^{\rm vac}$ and $\phi_{Q}^{\rm vac}$. Equations~(\ref{PhiqqGEq}),~(\ref{DeltaqqGEq}) and~(\ref{PiqqMass}) enable us to determine the behavior of $\phi_{q}$ in the superfluid phase as
\begin{eqnarray}
\phi_{q} = \frac{\phi_{q}^{\rm vac}\big(m_\pi^{\rm vac}\big)^2}{4\mu_{q}^2}\ . \label{Phiqq}
\end{eqnarray}
Thus, in a limit of $\mu_q\to\infty$, $\phi_{q}$ vanishes which indicates the restoration of chiral symmetry. Meanwhile, there is no concise expression of $\phi_{Q}$ in the superfluid phase. When the parameters satisfy $ 24c-\lambda_1=0$, the right-hand side (r.h.s.) of Eq.~(\ref{PhiQQGEq}) vanishes for which the $\phi_{Q}$ turns out to be always constant. In other words, the fate of $\phi_{Q}$ depends on the hierarchy of  $24c$ and $\lambda_1$. In order to visualize this structure we depict $\mu_q$ dependences of $\phi_{Q}$ in Fig.~\ref{fig:ThreeMFs}, together with those of $\phi_q$ and  $\Delta_{q}$ for completeness. In this figure the value of $c$ is changed with $\lambda_1=6$ being fixed as a demonstration.\footnote{The anomalous coupling $c$ must be positive to correctly reproduce $m_{\eta_{qq}}> m_{\pi_{qq}}$, and thus we have took $\lambda_1=6$ to demonstrate the sign inversion of $24c-\lambda_1$. We again stress that $c$ and $\lambda_1$ are small due to the large $N_c$ suppression. Inputs~(\ref{InputPhysical}) yields $\lambda_2={\cal O}(10^2)$.} In QC$_2$D with $2+2$ flavors there are no priori inputs since no lattice simulations have been done~\cite{Hands:2007uc,Murakami:2022lmq}. The remaining parameters are adjusted to reproduce the following physical quantities~\cite{ParticleDataGroup:2024cfk}:
\begin{eqnarray}
&&m^{\rm vac}_{\pi_{qq}} = 138\, {\rm MeV}\ , \ \ m^{\rm vac}_{K_{Qq}} = 494\, {\rm MeV}\ , \nonumber\\
&& f^{\rm vac}_{\pi_{qq}} = 92\, {\rm MeV}\ , \ \ f^{\rm vac}_{K_{Qq}} = 110\, {\rm MeV}\ , \label{InputPhysical}
\end{eqnarray}
in the vacuum, with $\phi_{q}= f_{\pi_{qq}}$ and $\phi_{Q} = 2f_{K_{Qq}}-f_{\pi_{qq}}$. Figure~\ref{fig:ThreeMFs} indeed shows that the fate of $\phi_{Q}$ in the baryon superfluid phase is controlled by the magnitude of anomaly effect $c$, while $\phi_q$ is uniquely determined by Eq.~(\ref{Phiqq}). 

One remarkable result in Fig.~\ref{fig:ThreeMFs} is the enhancement of $\phi_Q$ in the superfluid phase for larger $c$, which may contradict our naive expectation from the chiral-symmetry restoration. This enhancement can be understood by focusing on the flavor-mixing structures of $c$ and $\lambda_1$ terms. That is, such flavor-mixing effects (for instance displayed in Fig.~\ref{fig:Anomaly}) generate contributions proportional to purely light $\Delta_{q}^2+\phi_{q}^2$ to the stationary condition of heavy $\phi_{Q}$, which is indeed indicated in the r.h.s of Eq.~(\ref{PhiQQGEq}). When $24c-\lambda_1>0$ this r.h.s. plays a role of source term to create more $\phi_{Q}$, since in this case the left-hand side (l.h.s.) of Eq.~(\ref{PhiQQGEq}) must be magnified to compensate the r.h.s.. Moreover, the LSM analysis predicts that breaking strength of the Pauli-G\"{u}rsey symmetry measured by $\sqrt{\Delta_{q}^2+\phi_{q}^2}$ grows with $\mu_q$ in the superfluid phase, as explicitly shown in Fig.~\ref{fig:DeltaPhiq}. Those properties catalyze $\phi_{Q}$ in the superfluid phase. We note that a similar behavior was also found in quark-meson coupling model analysis in isospin QCD (QCD$_I$) matter with $N_f=2+1$~\cite{Kojo:2024sca}.

\begin{table}[t]
\begin{center}
  \begin{tabular}{c||ccccccc}  \hline\hline
 & $m_0^2 $ & $\lambda_1$ & $\lambda_2$ & $c$ & $\bar{c}m_q/2$ & $\bar{c}m_Q/2$ \\  \hline
Set I & $(230\, {\rm MeV})^2$ & $6$ & $99.3$ & $1$ & $(85.2\, {\rm MeV})^3$ & $ (264\, {\rm MeV})^3$ \\
Set II & $-(737\, {\rm MeV})^2$ & $6$ & $195$ & $0$ & $(85.2\, {\rm MeV})^3$ & $ (264\, {\rm MeV})^3$ \\
\hline \hline
 \end{tabular}
\caption{Parameters employed in the analysis in Sec.~\ref{sec:HadronMass}.}
\label{tab:Parameters}
\end{center}
\end{table}

\begin{table*}[t]
\begin{center}
  \begin{tabular}{c||ccccccccccc}  \hline\hline
 & $m^{\rm vac}_{\pi_{qq}}$ & $m^{\rm vac}_{a_{0,qq}}$ & $m^{\rm vac}_{\sigma_{l}}$ & $m^{\rm vac}_{\sigma_{h}}$ & $m^{\rm vac}_{\eta_{l}}$ & $m^{\rm vac}_{\eta_{h}}$ & $m^{\rm vac}_{K(\bar{K})}$ & $m^{\rm vac}_{\kappa(\bar{\kappa})}$ & $m^{\rm vac}_{\pi_{QQ}}$ & $m^{\rm vac}_{a_{0,QQ}}$ \\  \hline
Set I ($c=1$) & $138^*$ & $1107$ & $608$ & $1260$ & $493$ & $ 1172$ & $494^*$ & $1181$ & $637$ & $1275$ \\
Set II ($c=0$) & $138^*$ & $920$ & $965$ & $1489$ & $138$ & $ 637$ & $494^*$ & $1181$ & $637$ & $1416$ \\
\hline \hline
 \end{tabular}
\caption{Hadron masses in the vacuum estimated by the two parameter sets in Table~\ref{tab:Parameters}, in units of MeV. Note that there are mass degenerates between mesons and (anti)baryons due to the original Pauli-G\"{u}rsey $SU(8)$ symmetry, shown in Eq.~(\ref{PGSymmetryMass}) as well.}
\label{tab:VacuumMass}
\end{center}
\end{table*}

\subsection{Hadron mass spectrum at finite $\mu_q$}
\label{sec:HadronMass}

In this subsection we present numerical results on the hadron mass spectrum in cold QC$_2$D medium. Here we take $\mu_q=\mu_Q$.

As a demonstration we adopt the following two parameter sets listed in Table~\ref{tab:Parameters}, which are fixed by the inputs~(\ref{InputPhysical}) with ($c,\lambda_1)=(1,6)$ and $(0,6)$.\footnote{The sign of $m_0^2$ is positive for the parameter set I. The ``wine-bottle shape'' of the potential~(\ref{VLSM}) is driven by the anomalous coupling $c$.} The parameter set II corresponds to an anomaly-free system examined to gain insights into impacts of the anomaly effects. In the vacuum these parameters yield mass spectra tabulated in Table~\ref{tab:VacuumMass}. We note that the original Pauli-G\"{u}rsey symmetry imposes the following mass degeneracies:
\begin{eqnarray}
&& m^{\rm vac}_{\pi_{qq}} = m^{\rm vac}_{B_{qq}(\bar{B}_{qq})}\ , \ \  m^{\rm vac}_{a_{0,qq}} =  m^{\rm vac}_{B'_{qq}(\bar{B}'_{qq})} \ , \nonumber\\
 && m^{\rm vac}_{K_{Qq}(\bar{K}_{Qq})} =  m^{\rm vac}_{B_{Qq}(\bar{B}_{Qq})}  \ ,  \ \ m^{\rm vac}_{\kappa_{Qq}(\bar{\kappa}_{Qq})} =  m^{\rm vac}_{B'_{Qq}(\bar{B}'_{Qq})} \ , \nonumber\\
 &&  m^{\rm vac}_{\pi_{QQ}} = m^{\rm vac}_{B_{QQ}(\bar{B}_{QQ})}\ ,\ \  m^{\rm vac}_{a_{0,QQ}} =  m^{\rm vac}_{B'_{QQ}(\bar{B}'_{QQ})} \ . \nonumber\\ \label{PGSymmetryMass}
\end{eqnarray}
In Table~\ref{tab:VacuumMass}, $\sigma_l$ and $\sigma_h$ denote mass eigenstates in $\sigma_{qq}$ - $\sigma_{QQ}$ sector, the corresponding eigenvalues of which are smaller and larger, respectively. The same notation follows for $\eta_{l}$ and $\eta_{h}$. 

Depicted in the top panels of Figs.~\ref{fig:MassLLHH} -~\ref{fig:MassHH} are the resultant mass spectra with the parameter set I in Table~\ref{tab:Parameters}. The bottom ones are those with the parameter set II where the $U(1)$ axial anomaly effects are absent. We have displayed light, heavy-light and heavy hadrons separately in this order in the figures, where $\sigma_{h}$ and $\eta_h$ are contained in Fig.~\ref{fig:MassLLHH} due to the mixings. The vertical dotted lines denote the critical chemical potential $\mu_q^{\rm cr}$. In the superfluid phase, some mixings induced by the spontaneous breakdown of $U(1)$ light-quark number symmetry occur. On the other hand, $U(1)$ heavy-quark number symmetry is always protected for which no contaminations among light, heavy-light, and heavy sectors emerge.

The left panels of Fig.~\ref{fig:MassLLHH} indicate that a massless mode is obtained as the lowest excitation of $\sigma_{qq}$ - $\sigma_{QQ}$ - $B_{qq}$ - $\bar{B}_{qq}$ mixed state, in the superfluid phase. This zero mode is regarded as a Nambu-Goldstone (NG) boson associated with the $U(1)$ light-quark number violation, owning to the appearance of nonzero $\Delta_q$. Meanwhile, from the right panels of Fig.~\ref{fig:MassLLHH} one can find that the lowest mode of $\eta_{qq}$ - $\eta_{QQ}$ - $B_{qq}'$ - $\bar{B}_{qq}'$ mixed state, which would be dominated by $B_{qq}'$ component, gets lighter than $\pi_{qq}$ in the deeper superfluid phase; The light pion $\pi_{qq}$ is no longer the lightest excitation of $0^-$ hadron there. This is one of the noteworthy properties of the LSM analysis~\cite{Suenaga:2025sln} that cannot be captured by the ChPT framework~\cite{Kogut:1999iv,Kogut:2000ek}, and the lattice results indeed show this peculiar mass ordering~\cite{Murakami:2022lmq}. The light pion mass in the superfluid phase reads $m_{\pi_{\rm qq}} = 2\mu_q$ as obtained from Eq.~(\ref{DeltaCond}), which is universally derived from other chiral models as well~\cite{Suenaga:2025sln}. The lattice results with a heavy pion mass seem to be consistent with this formula within error bars~\cite{Hands:2007uc,Murakami:2022lmq}. It should be noted that our present findings are qualitatively consistent with the previous work based on the LSM with $N_f=2$~\cite{Suenaga:2022uqn,Suenaga:2023xwa}. 

The masses of $\eta_{l}$ and $\eta_{h}$ mesons are sensitive to the $U(1)$ axial anomaly effects as explicitly shown in Table~\ref{tab:VacuumMass}. In the absence of the anomaly effects, $\eta_{l}$ and $\eta_h$ masses are reduced to $\pi_{qq}$ and $\pi_{QQ}$ ones, respectively. Meanwhile, when the anomaly effects are sufficient $\eta_{l}$ and $\eta_{h}$ masses get heavier, particularly the $\eta_{h}$ mass is significantly enhanced due to the additional assist from $\eta_{qq}$ - $\eta_{QQ}$ mixing. As a result, in the vacuum $\eta_{h}$ can become heavier than $B_{qq}'$ ($\bar{B}_{qq}'$). As we access finite $\mu_q$, the  top-right panel of Fig.~\ref{fig:MassLLHH} indicates that a level crossing between $\eta_h$ and $\bar{B}'_{qq}$ occurs in the hadronic phase, due to the $\mu_q$ effects for $\bar{B}'_{qq}$ carrying $N_q=-2$.

Figure~\ref{fig:MassHL} indicates that the lowest modes of $\kappa_{Qq}$ - $\bar{\kappa}_{Qq}$ - $B_{Qq}$ - $\bar{B}_{Qq}$ and $K_{Qq}$ - $\bar{K}_{Qq}$ - $B'_{Qq}$ - $\bar{B}'_{Qq}$ mixed states in the superfluid phase monotonically decrease with $\mu_q$. Those reductions are again provided by the baryonic components $B_{Qq}$ or $B_{Qq}'$ essentially. Those masses approach zero but do not become negative even in a limit of $\mu_q\to\infty$, owning to the mixings. By comparing top and bottom panels, one can find that the $U(1)$ axial anomaly less affects $B_{Qq}$ ($\bar{B}_{Qq}$) and $B'_{Qq}$ ($B'_{Qq}$) masses, as long as the $K_{Qq}$ mass is employed as an input.

In the heavy sector exhibited in Fig.~\ref{fig:MassHH}, the $U(1)$ axial anomaly does not influence the mass spectrum significantly in the hadronic phase. Meanwhile, subtle variations of the mass modifications are seen in the superfluid phase. We note that in the heavy sector, there are no mixings among mesons and (anti)baryons from $U(1)$ heavy-quark number symmetry, and hence the monotonically decreasing $B_{QQ}$ and $B'_{QQ}$ masses indicated in Fig.~\ref{fig:MassHH} go across zero at certain $\mu_q$. In particular, the mass of $B_{QQ}$ would possibly hit zero at comparably small $\mu_q$, which results in the instability of $B_{QQ}$ and novel superfluidity violating $U(1)$ heavy-quark number conservation. For this reason we exhibit Fig.~\ref{fig:FateOfBQQ} focusing on the zero crossing of $B_{QQ}$ mass ($m_{B_{QQ}}$) to take a closer look at the fate of $B_{QQ}$ state. In plotting this figure we have only varied the anomalous coupling $c$, where $\lambda_1=6$ is fixed and the remaining parameters are adjusted to reproduce the inputs~(\ref{InputPhysical}). Figure~\ref{fig:FateOfBQQ} implies that the stronger anomaly effect $c$ leads to the zero crossing at smaller $\mu_q$. In this sense the anomaly effect catalyzes appearance of the heavy superfluidity.

Thus far,  QC$_2$D lattice simulations have been done with $N_f=2$, with which the pion mass is comparably large ($m^{\rm vac}_{\pi_{qq}} \gtrsim 700$ MeV)~\cite{Hands:2007uc,Murakami:2022lmq}. As supplemental demonstrations, we exhibit hadron mass spectra with such a large pion mass in Appendix~\ref{sec:HadronMassHeavyPion}.

\subsection{Topological susceptibility}
\label{sec:Topological}

Here we study the topological susceptibility to gain further insights into roles of the $U(1)$ axial anomaly and chiral-symmetry restoration in cold and dense QC$_2$D medium.

The topological susceptibility is defined by a space-time integrated two-point function of the gluon topological operator $Q=(g_s^2/64\pi^2)\epsilon^{\mu\nu\rho\sigma}G_{\mu\nu}^aG_{\rho\sigma}^a$~\cite{Astrakhantsev:2020tdl,Iida:2024irv}:
\begin{eqnarray}
\chi_{\rm top} = -\int d^4x\langle 0|{\rm T}Q(x)Q(0)|0\rangle\ ,
\end{eqnarray}
where ${\rm T}$ is the time-ordering operator. After a $U(1)$ axial transformation of the QC$_2$D partition function and adopting chiral Ward-Takahashi identities, this $\chi_{\rm top}$ can be expressed by Fermionic operators as\footnote{This equation is straightforwardly derived by extending the procedure shown in Ref.~\cite{Kawaguchi:2020qvg}.}
\begin{eqnarray}
\chi_{\rm top} = \chi^{qq}_{\rm top}  + \chi^{QQ}_{\rm top}  + \chi^{Qq}_{\rm top} \ ,
\end{eqnarray}
with
\begin{eqnarray}
\chi^{qq}_{\rm top} &=& i\bar{m}^2\left(\chi_{\pi_{qq}}  -\chi_{\eta_{qq}} \right) \ , \nonumber\\
\chi^{QQ}_{\rm top} &=& i\bar{m}^2\left(\chi_{\pi_{QQ}}  -\chi_{\eta_{QQ}} \right) \ , \nonumber\\
\chi^{Qq}_{\rm top} &=& -2i\bar{m}^2\chi_{\eta_{qq}\eta_{QQ}} \ . \label{ChiTopEach}
\end{eqnarray}
In these equations,
\begin{eqnarray}
\bar{m} = \left( 2/m_q+2/m_Q\right)^{-1}
\end{eqnarray}
is a quark mass parameter which is introduced to satisfy the flavor-singlet condition of the topological susceptibility~\cite{Baluni:1978rf}, and $\chi$'s are susceptibility functions of the corresponding operators ($\tau_f^a$ is the Pauli matrix in each flavor space)
\begin{eqnarray}
\chi_{\pi_{qq}}\delta^{ab} &=& \int d^4x\langle 0|{\rm T}(\bar{q}i\gamma_5\tau_f^aq)(x)(\bar{q}i\gamma_5\tau_f^bq)(0)|0\rangle\ , \nonumber\\
\chi_{\pi_{QQ}}\delta^{ab} &=& \int d^4x\langle 0|{\rm T}(\bar{Q}i\gamma_5\tau_f^aQ)(x)(\bar{Q}i\gamma_5\tau_f^bQ)(0)|0\rangle\ , \nonumber\\
\chi_{\eta_{qq}} &=& \int d^4x\langle 0|{\rm T}(\bar{q}i\gamma_5{\bm 1}_fq)(x)(\bar{q}i\gamma_5{\bm 1}_fq)(0)|0\rangle\ , \nonumber\\
\chi_{\eta_{qq}\eta_{QQ}} &=& \int d^4x\langle 0|{\rm T}(\bar{q}i\gamma_5{\bm 1}_fq)(x)(\bar{Q}i\gamma_5{\bm 1}_fQ)(0)|0\rangle\ , \nonumber\\
\chi_{\eta_{QQ}} &=& \int d^4x\langle 0|{\rm T}(\bar{Q}i\gamma_5{\bm 1}_fQ)(x)(\bar{Q}i\gamma_5{\bm 1}_fQ)(0)|0\rangle\ . \nonumber\\
\end{eqnarray}
One can easily find, e.g., $\bar{q}i\gamma_5\tau_f^aq = -\sqrt{2}\bar{c}{\cal P}^a$ ($a=1$ - $3$), by matching QC$_2$D with the present LSM~\cite{Suenaga:2025sln}, and thus those susceptibility functions are evaluate by
\begin{eqnarray}
&& \chi_{\pi_{qq}} = 2\bar{c}^2\frac{i}{-m_{\pi_{qq}}^2}\ , \ \ \chi_{\pi_{QQ}} = 2\bar{c}^2\frac{i}{-m_{\pi_{QQ}}^2}\ , \nonumber\\ 
&& \chi_{\eta_{qq}} = 2\bar{c}^2(i{\cal D}_{\eta B'})_{11}|_{p=0}\ , \ \ \chi_{\eta_{QQ}} = 2\bar{c}^2(i{\cal D}_{\eta B'})_{22}|_{p=0} \ , \nonumber\\
&& \chi_{\eta_{qq}\eta_{QQ}} = 2\bar{c}^2(i{\cal D}_{\eta B'})_{12}|_{p=0}\ , \label{ChiSus}
\end{eqnarray}
where the propagator matrix ${\cal D}_{\eta B'}$ is given in Eq.~(\ref{EtaBpPropagator}) in Appendix~\ref{sec:MassFormulas}. Substituting Eq.~(\ref{ChiSus}) into Eq.~(\ref{ChiTopEach}), the topological susceptibility $\chi_{\rm top}$ can be evaluated.

\begin{figure}[t]
\centering
\hspace*{-0.5cm} 
\includegraphics*[scale=0.3]{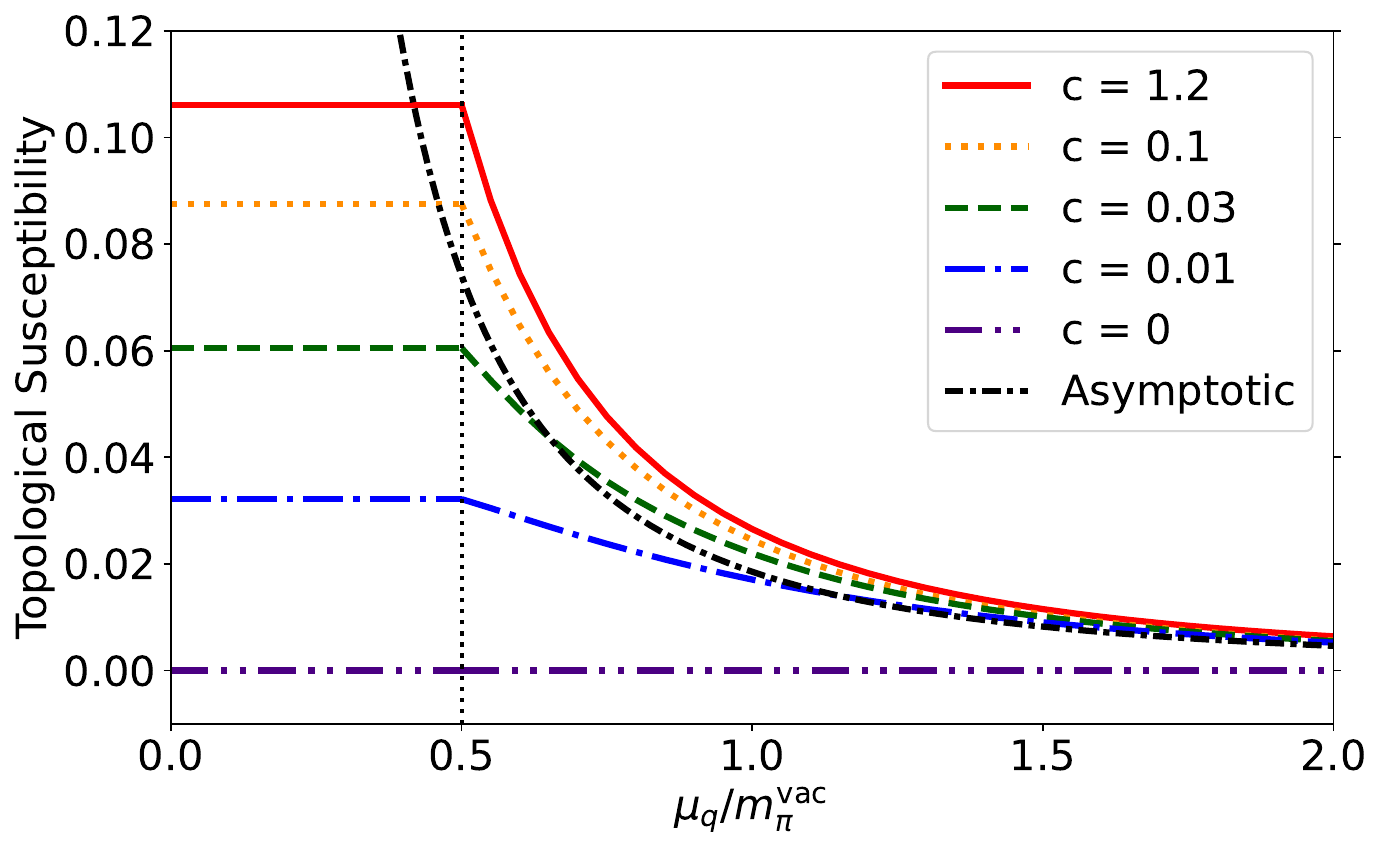}
\caption{$\mu_q$ dependences of the topological susceptibility $\chi_{\rm top}$ normalized by $(m_{\pi_{qq}}^{\rm vac})^4$, with several choices of $c$. The dashed curve is the asymptotic expression~(\ref{ChiAS}).}
\label{fig:ChiTop}
\end{figure}

Depicted in Fig.~\ref{fig:ChiTop} is the resultant $\mu_q$ dependences of the topological susceptibility $\chi_{\rm top}$ normalized by $(m_{\pi_{qq}}^{\rm vac})^4$, with several choices of $c$. The dashed curve exhibits an analytic formula that is expected to hold in the asymptotic region $\mu_q\gg m_\pi^{\rm vac}$ with $N_f=2$~\cite{Kawaguchi:2023olk}:
\begin{eqnarray}
\chi_{\rm top}^{\rm as.}/(m_{\pi_{qq}}^{\rm vac})^4 \sim \frac{(\phi^{\rm vac}_{q})^2}{24}\mu_q^{-2} \ . \label{ChiAS}
\end{eqnarray}
As discussed in Ref.~\cite{Kawaguchi:2023olk} in detail, the damping behavior proportional to $\mu_q^{-2}$ essentially originates from the chiral-symmetry restoration of $\phi_q\propto \mu_q^{-2}$ in the superfluid phase [see Eq.~(\ref{Phiqq})]. Figure~\ref{fig:ChiTop} indicates that the topological susceptibility is suppressed as $\mu_q$ is increased so as to follow the anomaly independent expression~(\ref{ChiAS}), while in the hadronic phase a larger $\chi_{\rm top}$ is generated by the sufficient anomaly effects. Therefore, regardless of the strength of the anomaly effects, the topological susceptibility is monotonically suppressed in dense regime followed by the chiral-symmetry restoration. This finding is consistent with the previous examination for $N_f=2$ case~\cite{Kawaguchi:2023olk,Fejos:2025oxi}. We note that $\chi_{\rm top}$ is always vanishing when the anomaly effects are switched off.

\begin{figure}[t]
\centering
\hspace*{-0.5cm} 
\includegraphics*[scale=0.3]{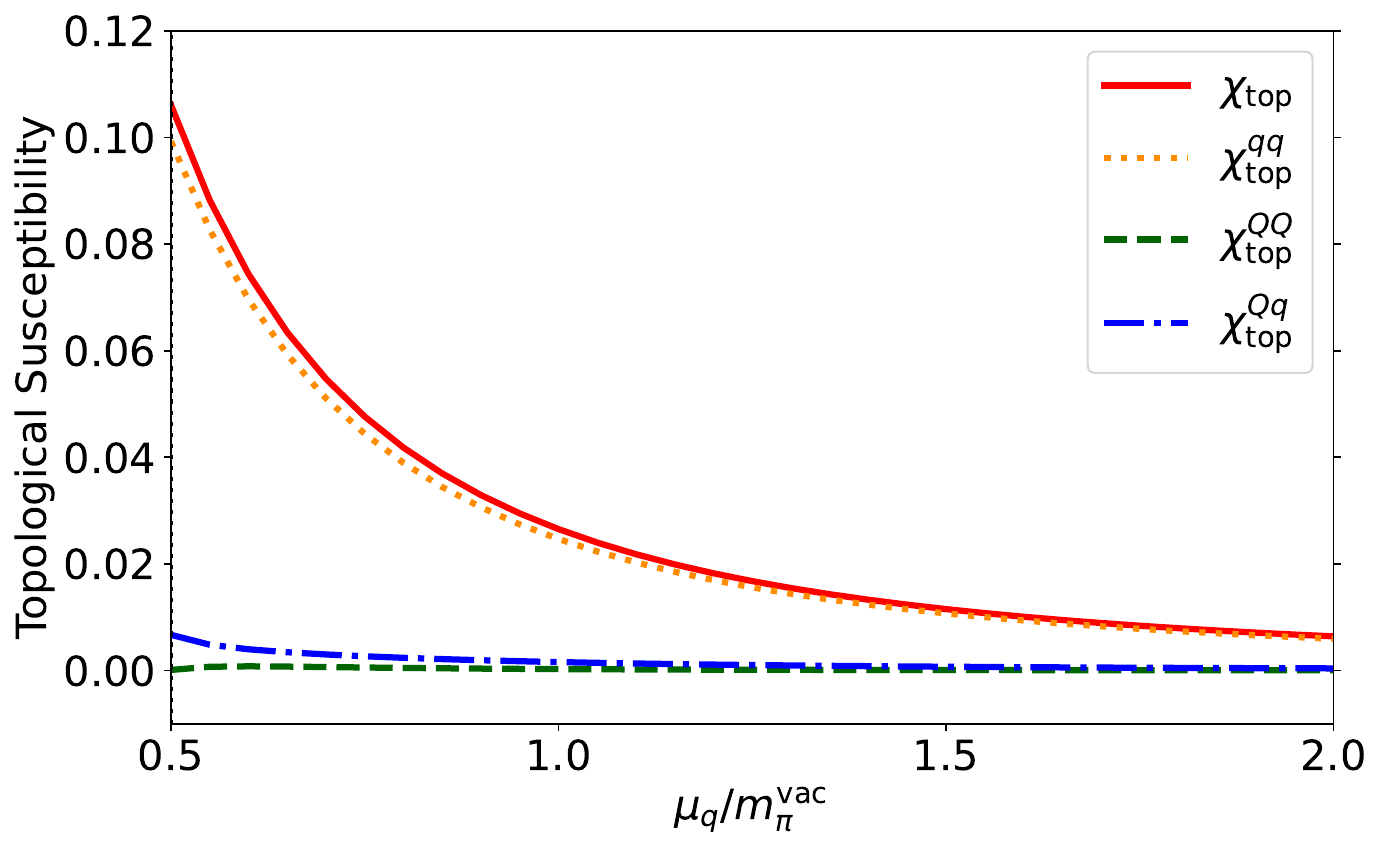}
\caption{$\mu_q$ dependences of each contribution $\chi_{\rm top}^{qq}$, $\chi_{\rm top}^{QQ}$ and $\chi_{\rm top}^{Qq}$ with $c=1.2$. The solid curve corresponds to the full result $\chi_{\rm top}$. The susceptibilities are normalized by $(m_{\pi_{qq}}^{\rm vac})^4$.}
\label{fig:ChiTopEach}
\end{figure}

In order to gain insights into the structure of the topological susceptibility, in Fig.~\ref{fig:ChiTopEach}, we plot $\chi_{\rm top}^{qq}$, $\chi_{\rm top}^{QQ}$ and $\chi_{\rm top}^{Qq}$ defined in Eq.~(\ref{ChiTopEach}) separately, with $c=1.2$. From this figure one can see that the fate of the topological susceptibility is governed by the light sector $\chi_{\rm top}^{qq}$. The suppression of $\chi_{\rm top}^{QQ}$ and $\chi_{\rm top}^{Qq}$ are simply understood by their heavy elements and persists to smaller $c$. We note that the predominant $\chi_{\rm top}^{qq}$ is mostly generated by $\chi_{\pi_{qq}}$ when we take larger $c$, that is, in the presence of the sufficient anomaly effects the topological susceptibility is solely governed by the light pion contribution.


\section{NJL analysis}
\label{sec:NJLAnalysis}

While at lower density hadronic picture is reasonably adopted, at higher density quark degrees of freedom cannot be ignored as naturally expected from the asymptotic freedom. The latter would not be captured by the present simple LSM analysis based on the mean-field approximation. Hence, in this section we employ the NJL model treating quarks explicitly to re-examine the increasing of $\phi_Q$ in the superfluid phase induced by the $U(1)$ axial anomaly effects, as the simplest demonstration~\cite{Hatsuda:1994pi,Buballa:2003qv}.

The NJL model based on the Pauli-G\"{u}rsey $SU(8)$ symmetry for $2+2$ flavors is of the form ($\mu_q=\mu_Q$)
\begin{eqnarray}
{\cal L}_{\rm NJL} &=& \bar{\psi}(i\Slash{\partial}+\mu_q\gamma_0-M)\psi+4G{\rm tr}[\Phi^\dagger\Phi] \nonumber\\
&& + \frac{K}{24}\Big( \epsilon_{ijklmnop}\Phi_{ij}\Phi_{kl}\Phi_{mn}\Phi_{op}  + {\rm H.c.}\Big)\, , \label{NJLLagrangian}
\end{eqnarray}
where $\psi=(q_1,q_2,Q_1,Q_2)^T$ is a four-component quark field and $M={\rm diag}(m_q,m_q,m_Q,m_Q)$ is a quark mass matrix. The $8\times8$ matrix
\begin{eqnarray}
\Phi_{ij} = \Psi_j^T\sigma^2\tau_c^2\Psi_i
\end{eqnarray}
is a quark composite operator with $\Psi = (\psi_R,\tilde{\psi}_L)$. In Eq.~(\ref{NJLLagrangian}), the $G$ term is corresponding to the familiar four-point interaction preserving $U(8)$ symmetry, whereas the last $K$ term is responsible for the $U(1)$ axial anomaly.

\begin{figure}[t]
\centering
\hspace*{-0.5cm} 
\includegraphics*[scale=0.3]{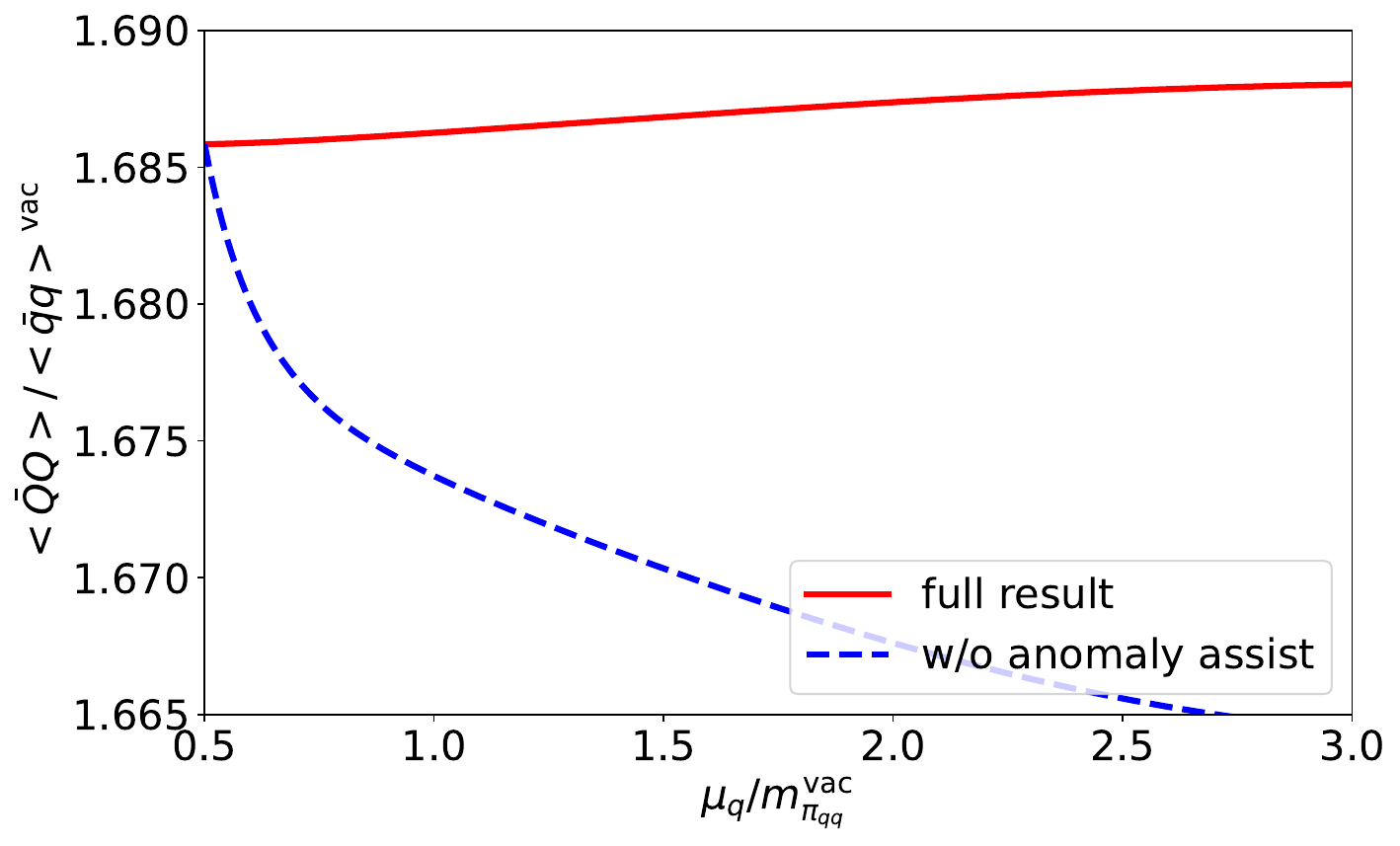}
\caption{$\mu_q$ dependences of the heavy chiral condensate $\langle\bar{Q}Q\rangle$. The dashed curve corresponds to a result with no anomaly assist.}
\label{fig:NJLQbarQ}
\end{figure}

We assume the following mean fields:
\begin{eqnarray}
&& \langle\bar{q}q\rangle \equiv \langle\bar{q}_1q_1\rangle = \langle\bar{q}_2q_2\rangle \ ,\ \ \langle qq\rangle \equiv  -\frac{1}{2}\langle q^TC\gamma_5\tau_c^2\tau_f^2q\rangle\ , \nonumber\\
&& \langle\bar{Q}Q\rangle \equiv \langle\bar{Q}_1Q_1\rangle = \langle\bar{Q}_2Q_2\rangle \ .
\end{eqnarray}
Making use of a quark one-loop approximation, gap equations for $\langle\bar{q}q\rangle$, $\langle\bar{Q}Q\rangle$ and $\langle qq\rangle$ are derived to be
\begin{eqnarray}
&&-8G\langle\bar{q}q\rangle - 12K\langle\bar{q}q\rangle\langle\bar{Q}Q\rangle^2 \nonumber\\
&=&  4(4G + 2K\langle\bar{Q}Q\rangle^2)\int_{\bm p}\frac{M_q}{E_{\bm p}^{(q)}}\left(\frac{\epsilon_{\rm p}^{(q)}({\bm p})}{\tilde{\epsilon}^{(q)}_{\rm p}({\bm p})} + \frac{\epsilon_{\rm a}^{(q)}({\bm p})}{\tilde{\epsilon}^{(q)}_{\rm a}({\bm p})}\right) \nonumber\\
&& + 32K\langle\bar{q}q\rangle\langle\bar{Q}Q\rangle\int_{\bm p}\theta(E_{\bm p}^{(Q)}-\mu)\frac{M_Q}{E_{\bm p}^{(Q)}}\ , \label{Gap1NJL}
\end{eqnarray}
\begin{eqnarray}
&&- 8G\langle\bar{Q}Q\rangle - 12K(\langle qq\rangle^2+\langle\bar{q}q\rangle^2)\langle\bar{Q}Q\rangle \nonumber\\
&=&  16K\langle qq\rangle \langle\bar{Q}Q\rangle \int_{\bm p}\left(\frac{\hat{\Delta}}{\tilde{\epsilon}^{(q)}_{\rm p}({\bm p})} + \frac{\hat{\Delta}}{\tilde{\epsilon}^{(q)}_{\rm a}({\bm p})}\right)  \nonumber\\
&& + 16K\langle\bar{q}q\rangle\langle\bar{Q}Q\rangle\int_{\bm p}\frac{M_q}{E_{\bm p}^{(q)}}\left(\frac{\epsilon_{\rm p}^{(q)}({\bm p})}{\tilde{\epsilon}^{(q)}_{\rm p}({\bm p})} + \frac{\epsilon_{\rm a}^{(q)}({\bm p})}{\tilde{\epsilon}^{(q)}_{\rm a}({\bm p})}\right) \nonumber\\
&& +8\big[4G + 2K(\langle qq\rangle^2+\langle\bar{q}q\rangle^2)\big]\int_{\bm p}\theta(E_{\bm p}^{(Q)}-\mu)\frac{M_Q}{E_{\bm p}^{(Q)}} \ ,\nonumber\\ \label{Gap2NJL}
\end{eqnarray}
and
\begin{eqnarray}
&& -8G\langle qq\rangle - 12K\langle qq\rangle\langle\bar{Q}Q\rangle^2 \nonumber\\
&=&  4(4G + 2K\langle\bar{Q}Q\rangle^2)\int_{\bm p}\left(\frac{\hat{\Delta}}{\tilde{\epsilon}^{(q)}_{\rm p}({\bm p})} + \frac{\hat{\Delta}}{\tilde{\epsilon}^{(q)}_{\rm a}({\bm p})}\right)    \nonumber\\
&& + 32K\langle qq\rangle \langle\bar{Q}Q\rangle\int_{\bm p}\theta(E_{\bm p}^{(Q)}-\mu)\frac{M_Q}{E_{\bm p}^{(Q)}}\ , \label{Gap3NJL}
\end{eqnarray}
respectively. In these equations we have defined dispersion relations by ($f=q,Q$)
\begin{eqnarray}
\epsilon_{\rm p}^{(f)}({\bm p}) &=& E^{(f)}_{\bm p}-\mu_q\ \  \nonumber\\
\epsilon_{\rm a}^{(f)}({\bm p}) &=& E^{(f)}_{\bm p} + \mu_q \ , \nonumber\\
\tilde{\epsilon}_{\rm p}^{(q)}({\bm p}) &=& \sqrt{(E_{\bm p}^{(q)}-\mu_q)^2 + \hat{\Delta}^2}\ ,\nonumber\\
\tilde{\epsilon}_{\rm a}^{(q)}({\bm p}) &=& \sqrt{(E_{\bm p}^{(q)}+\mu_q)^2 + \hat{\Delta}^2}\ ,
\end{eqnarray}
with $E_{\bm p}^{(f)} = \sqrt{{\bm p}^2 + M_f^2}$, where constituent quark masses read
\begin{eqnarray}
M_q &=& m_q-4G\langle\bar{q}q\rangle - 2K\langle\bar{q}q\rangle\langle\bar{Q}Q\rangle^2\ , \nonumber\\
M_Q &=& m_Q-4G\langle\bar{Q}Q\rangle - 2K(\langle qq\rangle^2+\langle\bar{q}q\rangle^2)\langle\bar{Q}Q\rangle \ ,\nonumber\\
\hat{\Delta} &=& (-4G- 2K\langle\bar{Q}Q\rangle^2)\langle qq\rangle\ .
\end{eqnarray}
Flavor-mixing roles of the $U(1)$ axial anomaly are clearly understood in these equations, for instance, $Q$-fluctuation contributions enter the gap equation for $\langle\bar{q}q\rangle$ via nonzero $K$ in Eq.~(\ref{Gap1NJL}). The detailed derivation is provided in Appendix.~\ref{sec:NJLDetail}. It should be noted that $\langle\bar{q}q\rangle$ and $\langle\bar{Q}Q\rangle$ are negative.

Solving the gap equations~(\ref{Gap1NJL}) -~(\ref{Gap3NJL}), $\mu_q$ dependences of the heavy chiral condensate $\langle\bar{Q}Q\rangle$ in the superfluid phase are evaluated, as in Fig.~\ref{fig:NJLQbarQ}. Here, as a demonstration we have chosen $m_q=5$ MeV, $m_Q=135$ MeV, $G\Lambda^2 = 2.8$ and $K\Lambda^8 = 15$, with $\Lambda=650$ MeV being a three-dimensional cutoff. These parameters yield $m_{\pi_{qq}}^{\rm vac} \sim138$ MeV and $m_{K_{Qq}}^{\rm vac} \sim 497$ MeV which are close to the physical values. The figure indicates that the heavy chiral condensate $\langle\bar{Q}Q\rangle$ is enhanced as $\mu_q$ increases, consistently with the LSM analysis. Therefore, in the superfluidity we can conclude that the heavy chiral condensate is enhanced even when the quark degrees of freedom governs matter in cold-dense QC$_2$D with $N_f=2+2$, provided that the adequate anomaly effects exist. It should be noted that we have plotted the figure for only $\mu_q\geq m_{\pi_{qq}}^{\rm vac}/2$, since below this critical chemical potential the condensates are constant, similarly to the LSM results in Fig.~\ref{fig:ThreeMFs}.

Within the NJL framework, growth of the mean fields are determined by competitions between the tree-level contributions and quantum fluctuations: l.h.s. and r.h.s. of Eqs.~(\ref{Gap1NJL}) -~(\ref{Gap3NJL}) respectively, and hence the mean fields are magnified when many fluctuations contribute. Focusing on Eq.~(\ref{Gap2NJL}), one can find additional fluctuations proportional to $K\langle qq\rangle\hat{\Delta}$ in the first line of r.h.s., which could be regarded as dominant anomaly-assisted $q$ contributions with the diquark condensate $\langle qq\rangle$. In order to quantify these contributions we have also exhibit $\mu_q$ dependences of $\langle\bar{Q}Q\rangle$ evaluated without this term in Fig.~\ref{fig:NJLQbarQ} (dashed), which is suppressed with $\mu_q$. Therefore one can understand that the enhancement is triggered by the anomaly effects with creation of the superfluidity. We note that when $\mu_q$ becomes sufficiently large, the Fermi sea of $Q$ is created for which the $Q$ fluctuations are directly prevented, and finally the $\langle\bar{Q}Q\rangle$ would turn to be suppressed.


\begin{figure}[t]
\centering
\hspace*{-0.5cm} 
\includegraphics*[scale=0.3]{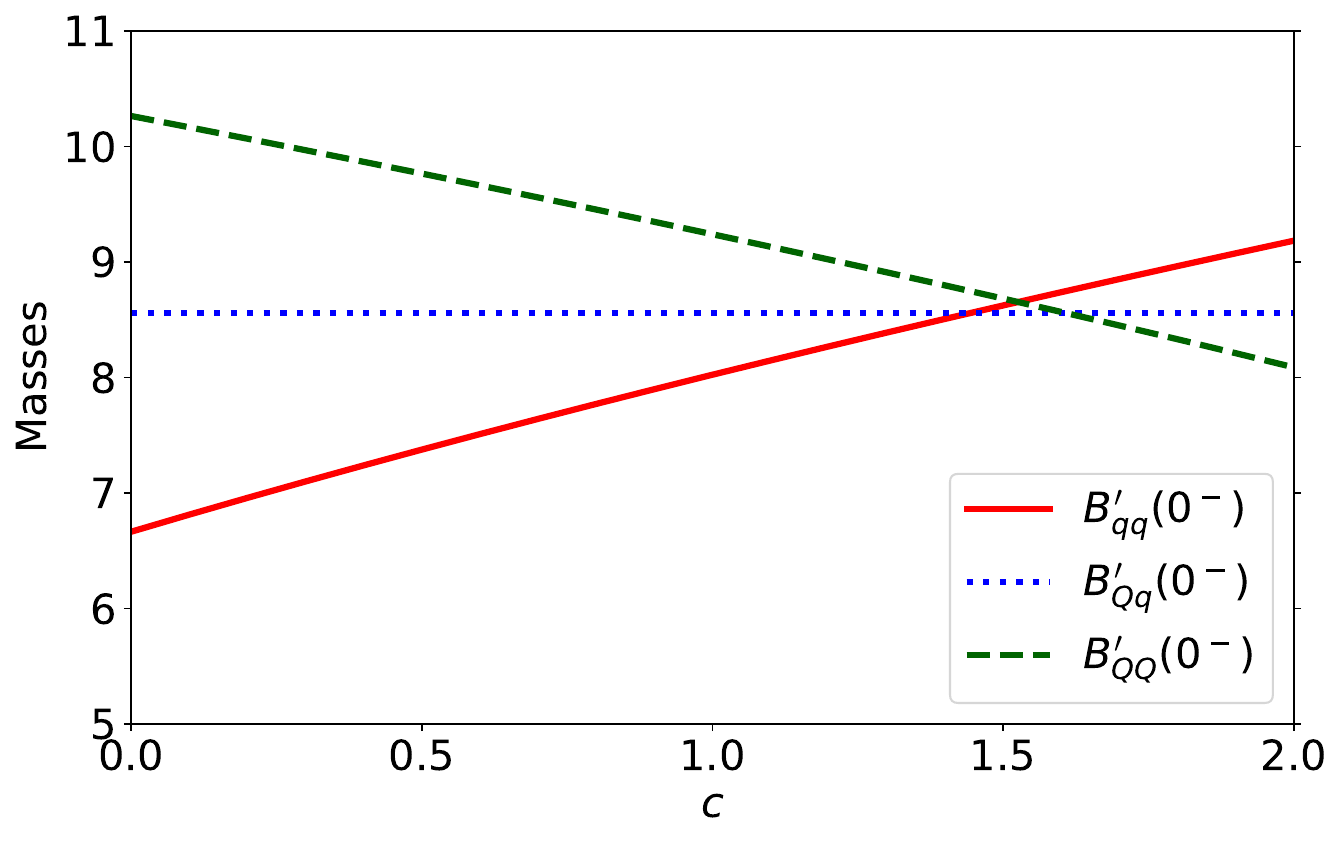}
\caption{$c$ dependences of $B'_{qq}$, $B'_{Qq}$ and $B'_{QQ}$ masses in the vacuum. Those masses are normalized by $m_{\pi_{qq}}^{\rm vac}$.}
\label{fig:InverseMH}
\end{figure}

\section{Complete inverse mass hierarchy: ${m}_{B'_{QQ}}^{\rm vac}<{m}_{B'_{Qq}}^{\rm vac}<{m}_{B'_{qq}}^{\rm vac}$}
\label{sec:InverseMH}

From effective-model analyses in $N_c=3$ based on the linear representation of chiral symmetry, it is predicted that negative-parity diquarks show the so-called {\it inverse mass hierarchy} where $[qq]_0^-$ diquark would become heavier than $[sq]_0^-$ one: $m^{\rm vac}_{[qq]_0^-}> m^{\rm vac}_{[sq]_0^-}$, provided the $U(1)$ axial anomaly effects are sufficient, despite their quark contents~\cite{Harada:2019udr,Suenaga:2023tcy,Suenaga:2024vwr}. In our $N_c=3$ world, diquarks are not direct observables due to the confinement, and their dynamics are reflected by SHBs, composed of one heavy quark and one (light) diquark, by virtue of the heavy-quark effective theory~\cite{manohar2000heavy}. Although the inverse mass hierarchy is regarded as one of the remarkable consequences of the $U(1)$ axial anomaly, the SHBs containing $[qq]_0^-$ or $[sq]_0^-$ have not been experimentally observed. Hence, QC$_2$D world where diquarks are well defined as hadrons can be an alternative testing ground to see this peculiar mass hierarchy. For these reasonings, here we focus on mass orderings of the negative-parity diquarks from the viewpoints of the anomaly effects.

In the present LSM, according to Appendix~\ref{sec:MassFormulas}, $c$ dependences of $B'_{qq}$, $B'_{Qq}$ and $B'_{QQ}$ masses in the vacuum ($\mu_q=0$) read
\begin{eqnarray}
&& ({m}_{B'_{qq}}^{\rm vac})^2 = ({m}_{B'_{qq}}^{\rm vac})^2|_{c=0} + 48c\big[(\phi_Q^{\rm vac})^2 - (\phi_q^{\rm vac})^2\big] \ ,\nonumber\\
&& ({m}_{B'_{Qq}}^{\rm vac})^2 = ({m}_{B'_{Qq}}^{\rm vac})^2|_{c=0} \ , \nonumber\\
&& ({m}_{B'_{QQ}}^{\rm vac})^2 = ({m}_{B'_{qq}}^{\rm vac})^2|_{c=0} - 48c\big[(\phi_Q^{\rm vac})^2 - (\phi_q^{\rm vac})^2\big] \ , \nonumber\\
\end{eqnarray}
respectively. Since the anomalous coupling $c$ is positive and $\phi_Q^{\rm vac}>\phi_q^{\rm vac}$, our LSM predicts increment and decrement of ${m}_{B'_{qq}}^{\rm vac}$ and ${m}_{B'_{QQ}}^{\rm vac}$, respectively, as the anomaly effect is enhanced, while  ${m}_{B'_{Qq}}^{\rm vac}$ is independent of $c$. To visualize these tendencies, in Fig.~\ref{fig:InverseMH} we depict $c$ dependences of vacuum masses of the $0^-$ diquarks with the inputs~(\ref{InputPhysical}). This figure indicates that the mass hierarchy of ${m}_{B'_{qq}}^{\rm vac}<{m}_{B'_{Qq}}^{\rm vac}<{m}_{B'_{QQ}}^{\rm vac}$ is inverted to be  
\begin{eqnarray}
{m}_{B'_{QQ}}^{\rm vac}<{m}_{B'_{Qq}}^{\rm vac}<{m}_{B'_{qq}}^{\rm vac}\ ,
\end{eqnarray}
 for $c\gtrsim 1.5$. Thus, our model predicts the complete inverse mass hierarchy of $0^-$ diquarks driven by sufficient anomaly effects with respect to their quark contents. Meanwhile, all the $0^+$ masses do not depend on $c$.
We note that all the baryon masses in the vacuum are independent of $\lambda_1$ as long as the inputs~(\ref{InputPhysical}) are adopted, as explained in Appendix~\ref{sec:MassFormulas} [See Eqs.~(\ref{MPiVac}) -~(\ref{MBQQVac})]. We also note that $c$ can be chosen to be up to $c\sim 2$ when changing the value of $\lambda_1$ accordingly, from which the potential $V_{\rm LSM}$ is certainly bounded from below.


\section{Conclusions}
\label{sec:Conclusions}

In this paper we have studied flavor-symmetry violation effects particularly focusing on the heavy chiral condensate $\langle\bar{Q}Q\rangle$ and hadron mass spectrum in cold and dense QC$_2$D, by means of the $N_f=2+2$ linear sigma model based on the approximate $SU(8)$ Pauli-G\"{u}rsey symmetry. We have paid particular attention to roles of the $U(1)$ axial anomaly which triggers the flavor-mixing structures of hadrons.

Adopting a mean-field approximation, we have found that the $U(1)$ axial anomaly effects are capable of enhancing the heavy chiral condensate $\langle\bar{Q}Q\rangle$ in the baryon superfluid phase where light diquark baryons $(\sim qq)$ condense. This enhancement is catalyzed by growth of breaking strength of the Pauli-G\"{u}rsey symmetry ($\sim\sqrt{\langle qq\rangle^2+\langle\bar{q}q\rangle^2}$) in the superfluid phase. The NJL model describing quark degrees of freedom explicitly, which play significant roles in more dense regime, has also derived the $\langle\bar{Q}Q\rangle$ enhancement in the superfluid phase, in the presence of the anomaly effects.

In the superfluid phase, any modes of the heavy-light hadrons do not become exactly massless owning to mixings triggered by the $U(1)$ light-quark number violation. Meanwhile, the heavy diquark baryon ($\sim QQ$) mass has shown a zero crossing at certain chemical potential. That is, a novel superfluidity filled by Bose-Einstein condensates of the heavy diquarks is predicted. We have found that stronger $U(1)$ axial anomaly effects enable this heavy-diquark superfluidity to emerge at lower $\mu_q$.

We have also investigated the topological susceptibility with $2+2$ flavors. Our results indicate that the susceptibility is suppressed in dense regime followed by the chiral-symmetry restoration, regardless of the strength of the anomaly effects. This anomaly-independent suppression is consistent with the previous analysis with $N_f=2$~\cite{Kawaguchi:2023olk,Fejos:2025oxi}.

Although our present LSM does not incorporate quark degrees of freedom explicitly, at larger $\mu_q$ the latter cannot be ignored. In particular, a dual picture between hadrons and quarks inside them based on the so-called {\it quark saturation} predicts minor roles of heavy hadrons in dense medium~\cite{Fujimoto:2023mzy,Fujimoto:2024doc}. Hence, for more precise determination of the onset $\mu_q$ of the heavy-diquark superfluidity, it is necessary to take into account quarks~\cite{Chiba:2023ftg,Chiba:2024cny,Kojo:2024sca}. We leave such examinations for future study.

In addition to the analysis in dense regime, we have studied a mass ordering of the negative-parity diquarks focusing on the anomaly effects. As a result it is found that the sufficiently large anomaly effects could lead to a complete inverse mass hierarchy where $m_{QQ(0^-)}<m_{Qq(0^-)}<m_{qq(0^-)}$, despite naive expectations from their quark contents.

Our findings are expected to provide future QC$_2$D lattice simulations with useful information on flavor-symmetry violation, particularly from viewpoints of the $U(1)$ axial anomaly, in cold and dense medium.

\section*{Acknowledgment}

M.~S. was supported by Grant-in-Aid for JSPS Fellows No.~25KJ1433. D.~S. was supported by Grants-in-Aid for Scientific Research No.~23K03377, No.~23H05439 and No.~25K17386, from JSPS. D.~S. thanks Toru Kojo for fruitful discussions on the role of quarks inside hadrons, and Mamiya Kawaguchi for comments on the topological susceptibility in dense medium. 

\appendix

\section{$SU(8)$ generators}
\label{sec:Generators}

The hadron field $\Sigma$ for $N_f=2+2$ QC$_2$D, defined in Eq.~(\ref{SigmaBilinear}), takes a form of $8\times 8$ matrix since originally the theory possesses the Pauli-G\"{u}rsey $SU(8)$ symmetry. Due to its antisymmetric property $\Sigma$ has complex $28$ degrees of freedom. In the present work we adopt the following $27$ generators of $G/H=SU(8)/Sp(8)$ to parametrize some of them:
\begin{eqnarray}
&& X^{a=1-15} = \frac{1}{2\sqrt{2}}\left(
\begin{array}{cc}
\eta_f^a & 0 \\
0 &(\eta_f^a)^T \\
\end{array}
\right)_{8\times 8}\ , 
\end{eqnarray}
and
\begin{eqnarray}
&&  X^{a=16-27} = \frac{1}{2\sqrt{2}}\left(
\begin{array}{cc}
0& D_f^a \\
(D_f^a)^\dagger & 0 \\
\end{array}
\right)_{8\times8} \ .
\end{eqnarray}
In these equations we have defined $4\times4$ matrices
\begin{eqnarray}
&&  D_f^{16}=\eta_f^2\ , \ \ D_f^{17}=i\eta_f^2\ , \ \ D_f^{18}=\eta_f^5\ , \ \ D_f^{19}=i\eta_f^5\ , \nonumber\\
&& D_f^{20}=\eta_f^7\ , \ \ D_f^{21}=i\eta_f^7\ , \ \  D_f^{22}=\eta_f^{10}\ , \ \ D_f^{23}=i\eta_f^{10}\ , \nonumber\\
 && D_f^{24}=\eta_f^{12}\ , \ \ D_f^{25}=i\eta_f^{12}\ , \ \  D_f^{26}=\eta_f^{14}\ , \ \ D_f^{27}=i\eta_f^{14}\ , \nonumber\\
\end{eqnarray}
where
\begin{eqnarray}
\eta_f^{a} = \left(
\begin{array}{cc}
\lambda_f^a & 0 \\
0 & 0 \\
\end{array}
\right)_{4\times4} \ \ \ \ (a=1-8)
\end{eqnarray}
with $\lambda_f^a$ being the $3\times3$ Gell-Mann matrices, and
\begin{eqnarray}
&& \eta_f^9 =  \left(
\begin{array}{cccc}
0 & 0 & 0 & 1 \\
0 & 0 & 0 & 0 \\
0 & 0 & 0 & 0 \\
1 & 0 & 0 & 0 \\
\end{array}
\right) \ , \ \ \eta^{10}_f =  \left(
\begin{array}{cccc}
0 & 0 & 0 & -i \\
0 & 0 & 0 & 0 \\
0 & 0 & 0 & 0 \\
i & 0 & 0 & 0 \\
\end{array}
\right) \ , \nonumber\\
&& \eta^{11}_f =  \left(
\begin{array}{cccc}
0 & 0 & 0 & 0 \\
0 & 0 & 0 & 1 \\
0 & 0 & 0 & 0 \\
0 & 1 & 0 & 0 \\
\end{array}
\right) \ , \ \  \eta^{12}_f =  \left(
\begin{array}{cccc}
0 & 0 & 0 & 0 \\
0 & 0 & 0 & -i \\
0 & 0 & 0 & 0 \\
0 & i & 0 & 0 \\
\end{array}
\right)\ , \nonumber\\
&&  \eta^{13}_f =  \left(
\begin{array}{cccc}
0 & 0 & 0 & 0 \\
0 & 0 & 0 & 0 \\
0 & 0 & 0 & 1 \\
0 & 0 & 1 & 0 \\
\end{array}
\right)\ , \ \  \eta^{14}_f =  \left(
\begin{array}{cccc}
0 & 0 & 0 & 0 \\
0 & 0 & 0 & 0 \\
0 & 0 & 0 & -i \\
0 & 0 & i & 0 \\
\end{array}
\right)\ , \nonumber\\
&& \eta^{15}_f =  \frac{1}{\sqrt{6}}\left(
\begin{array}{cccc}
1 & 0 & 0 & 0 \\
0 & 1 & 0 & 0 \\
0 & 0 & 1 & 0 \\
0 & 0 & 0 & -3 \\
\end{array}
\right)\ .
\end{eqnarray}
The remaining one degree of freedom is parametrized via
\begin{eqnarray}
X^{a=0} = \frac{1}{4}{\bm 1}_{8\times8}\ ,
\end{eqnarray}
belonging to the trivial algebra. These matrices satisfy
\begin{eqnarray}
&& {\rm tr}[X^a X^b] = \frac{\delta^{ab}}{2}\ , \nonumber\\
&& X^a {\bm E} = {\bm E}(X^a)^T\ , \label{XProperty}
\end{eqnarray}
with the $8\times 8$ symplectic matrix ${\bm E}$. Then, the matrix $\Sigma$ is given by Eq.~(\ref{SigmaDef}).

\section{Mass formulas}
\label{sec:MassFormulas}

Here we exhibit mass formulas of the hadrons at finite $\mu_q$ and $\mu_Q$ derived in our LSM~(\ref{LSM8}) at mean-filed level, with Eqs.~(\ref{PhiMean}) and~(\ref{DeltaMean}). 

The hadron masses are read off by pole positions of the corresponding propagator at the rest frame, which would take matrix forms owing to (kinetic) mixings. Among the hadrons, $\pi_{qq}$, $\pi_{QQ}$, $a_{0,qq}$ and $a_{0,QQ}$ are not contaminated by the mixings for which their masses are simply derived to be
\begin{eqnarray}
m_{\pi_{qq}}^2 &=& m_0^2+\lambda_1\left(\Delta_{q}^2+\phi_{q}^2+\phi_{Q}^2\right)  \nonumber\\
&& + \frac{\lambda_2}{4}\left(\Delta_{q}^2 + \phi_{q}^2\right) -24c\, \phi_{Q}^2 \ ,\label{PiqqMass}
\end{eqnarray}
\begin{eqnarray}
m_{\pi_{QQ}}^2 &=& m_0^2+\lambda_1\left(\Delta_{q}^2+\phi_{q}^2+\phi_{Q}^2\right) \nonumber\\
&& + \frac{\lambda_2}{4}\phi_{Q}^2 -24c\, (\Delta_{q}^2+\phi_{q}^2) \ ,\label{PiQQMass}
\end{eqnarray}
\begin{eqnarray}
m_{a_{0,qq}}^2 &=& m_0^2+\lambda_1\left(\Delta_{q}^2+\phi_{q}^2+\phi_{Q}^2\right) \nonumber\\
&& + \frac{3\lambda_2}{4}\left(\Delta_{q}^2 + \phi_{q}^2\right) +24c\, \phi_{Q}^2\ ,
\end{eqnarray}
and
\begin{eqnarray}
m_{a_{0,QQ}}^2 &=& m_0^2+\lambda_1\left(\Delta_q^2+\phi_q^2+\phi_Q^2\right)   \nonumber\\
&& + \frac{3\lambda_2}{4}\phi_Q^2 + 24c\, (\Delta_q^2 + \phi_q^2)\ ,
\end{eqnarray}
respectively, from the quadratic terms.

Since diquark condensates are generated by only light quarks, $U(1)$ heavy-quark number symmetry is preserved. Thus, the heavy (anti)diquarks, $(B_{QQ},\bar{B}_{QQ})$ and $(B_{QQ}',\bar{B}_{QQ}')$, are always separated by other hadrons. Those propagator-inverse matrices read
\begin{eqnarray}
{\cal D}_{B_{QQ}}^{-1} \equiv \left(
\begin{array}{cc}
p^2-m_{{\cal P}^{26}}^2 & 4i\mu_Q p_0 \\
-4i\mu_Q p_0 & p^2-m_{{\cal P}^{27}} \\
\end{array}
\right)\ ,
\end{eqnarray}
with
\begin{eqnarray}
m_{{\cal P}^{26}}^2 &=& -4\mu_{QQ}^2 + m_0^2+\lambda_1\left(\Delta_q^2+\phi_q^2+\phi_Q^2\right) \nonumber\\
&& + \frac{\lambda_2}{4}\phi_Q^2 - 24c\left(\phi_q^2 + \Delta_q^2\right)\ , \nonumber\\
m_{{\cal P}^{27}}^2 &=& -4\mu_{QQ}^2 + m_0^2+\lambda_1\left(\Delta_q^2+\phi_q^2+\phi_Q^2\right) \nonumber\\
&& + \frac{\lambda_2}{4}\phi_Q^2 - 24c\left(\phi_q^2 + \Delta_q^2\right)\ ,
\end{eqnarray}
and
\begin{eqnarray}
{\cal D}_{B'_{QQ}}^{-1} \equiv \left(
\begin{array}{cc}
p^2-m_{{\cal S}^{26}}^2 & 4i\mu_Q p_0 \\
-4i\mu_Q p_0 & p^2-m_{{\cal S}^{27}} \\
\end{array}
\right)\ ,
\end{eqnarray}
with
\begin{eqnarray}
m_{{\cal S}^{26}}^2 &=& -4\mu_{QQ}^2 + m_0^2+\lambda_1\left(\Delta_q^2+\phi_q^2+\phi_Q^2\right) \nonumber\\
&& + \frac{3\lambda_2}{4}\phi_Q^2 + 24c\left(\phi_q^2 + \Delta_q^2\right)\ , \nonumber\\
m_{{\cal S}^{27}}^2 &=& -4\mu_{QQ}^2 + m_0^2+\lambda_1\left(\Delta_q^2+\phi_q^2+\phi_Q^2\right) \nonumber\\
&& + \frac{3\lambda_2}{4}\phi_Q^2 + 24c\left(\phi_q^2 + \Delta_q^2\right)\ .
\end{eqnarray}

As shown in Eq.~(\ref{MixingStructure}), mixing structures appear within $(\eta_{qq},\eta_{QQ}, B'_{qq},\bar{B}'_{qq})$, $(\sigma_{qq},\sigma_{QQ}, B_{qq},\bar{B}_{qq})$, $(K_{Qq},\bar{K}_{Qq}, B'_{Qq},\bar{B}'_{Qq})$, and $(\kappa_{Qq},\bar{\kappa}_{Qq}, B_{Qq},\bar{B}_{Qq})$ sectors. The propagator-inverse matrix for $(\eta_{qq},\eta_{QQ}, B'_{qq},\bar{B}'_{qq})$ in momentum space reads
\begin{widetext}
\begin{eqnarray}
{\cal D}_{\eta B'}^{-1} \equiv \left(
\begin{array}{cccc}
p^2-m_{\eta_{qq}}^2  & -m_{\eta_{qq}\eta_{QQ}}^2 & 0 & -m_{\eta_{qq}{\cal S}^{17}}^2 \\
 -m_{\eta_{qq}\eta_{QQ}}^2 & p^2-m_{\eta_{QQ}}^2 & 0 & -m_{\eta_{QQ}{\cal S}^{17}}^2 \\
 0 & 0 & p^2-m_{{\cal S}^{16}}^2 & 4i\mu_q p_0 \\
 -m_{\eta_{qq}{\cal S}^{17}}^2 & -m_{\eta_{QQ}{\cal S}^{17}}^2 & -4i\mu_q p_0 & p^2-m_{{\cal S}^{17}}^2 \\
\end{array}
\right)\ , \label{EtaBpPropagator}
\end{eqnarray}
where
\begin{eqnarray}
&& m_{\eta_{qq}}^2 = m_0^2+\lambda_1\left(\Delta_q^2+\phi_q^2+\phi_Q^2\right) + \frac{\lambda_2}{4}\left(3\Delta_q^2 + \phi_q^2\right)+ 24c\, \phi_Q^2 \ , \ \  m_{\eta_{qq}\eta_{QQ}}^2  = 48c\phi_q\phi_Q\ ,\nonumber\\
&& m_{\eta_{QQ}}^2 = m_0^2+\lambda_1\left(\Delta_q^2+\phi_q^2+\phi_Q^2\right) + \frac{\lambda_2}{4}\phi_Q^2 + 24c\left(\phi_q^2+ \Delta_q^2\right)\ , \ \  m_{\eta_{qq}{\cal S}^{17}}^2 = \frac{\lambda_2}{2}\Delta_q\phi_q\ , \nonumber\\
&&m_{{\cal S}^{16}}^2 = -4\mu_{q}^2 + m_0^2+\lambda_1\left(\Delta_q^2+\phi_q^2+\phi_Q^2\right) + \frac{3\lambda_2}{4}\left(\Delta_q^2 + \phi_q^2\right) +24c\, \phi_Q^2\ , \ \  m_{\eta_{QQ}{\cal S}^{17}}^2  = -48c\Delta_q\phi_Q \ , \nonumber\\
&&m_{{\cal S}^{17}}^2 = -4\mu_{q}^2 + m_0^2+\lambda_1\left(\Delta_q^2+\phi_q^2+\phi_Q^2\right) + \frac{\lambda_2}{4}\left(\Delta_q^2 + 3\phi_q^2\right) + 24c\, \phi_Q^2 \ .
\end{eqnarray}
Then the masses are evaluated by solving ${\rm det}({\cal D}_{\eta B'}^{-1})=0$ with respect to $p_0$ with ${\bm p}={\bm 0}$.

The propagator-inverse matrix for $(\sigma_{qq},\sigma_{QQ}, B_{qq},\bar{B}_{qq})$ in momentum space reads
\begin{eqnarray}
{\cal D}_{\sigma B}^{-1} \equiv \left(
\begin{array}{cccc}
p^2-m_{\sigma_{qq}}^2  & -m_{\sigma_{qq}\sigma_{QQ}}^2 & 0 & -m_{\sigma_{qq}{\cal P}^{17}}^2 \\
 -m_{\sigma_{qq}\sigma_{QQ}}^2 & p^2-m_{\sigma_QQ}^2 & 0 & -m_{\sigma_{QQ}{\cal P}^{17}}^2 \\
 0 & 0 & p^2-m_{{\cal P}^{16}}^2 & 4i\mu_q p_0 \\
 -m_{\sigma_{qq}{\cal P}^{17}}^2 & -m_{\sigma_{QQ}{\cal P}^{17}}^2 & -4i\mu_q p_0 & p^2-m_{{\cal P}^{17}}^2 \\
\end{array}
\right)\ ,
\end{eqnarray}
where
\begin{eqnarray}
&& m_{\sigma_{qq}}^2 = m_0^2+\lambda_1\left(\Delta_q^2+3\phi_q^2+\phi_Q^2\right) + \frac{\lambda_2}{4}\left(\Delta_q^2 + 3\phi_q^2\right) - 24c\, \phi_Q^2\ , \ \  m_{\sigma_{qq}\sigma_{QQ}}^2  = 2\lambda_1\phi_q\phi_Q - 48c\phi_q\phi_Q\ , \nonumber\\
&& m_{\sigma_{QQ}}^2 = m_0^2+\lambda_1\left(\Delta_q^2+\phi_q^2+3\phi_Q^2\right) + \frac{3\lambda_2}{4}\phi_Q^2 - 24c\left(\phi_q^2+ \Delta_q^2\right)\ , \ \  m_{\sigma_{qq}{\cal P}^{17}}^2 = 2\lambda_1\Delta_q\phi_q + \frac{\lambda_2}{2}\Delta_q\phi_q \ ,\nonumber\\
&& m_{{\cal P}^{16}}^2 = -4\mu_{q}^2 + m_0^2+\lambda_1\left(\Delta_q^2+\phi_q^2+\phi_Q^2\right) + \frac{\lambda_2}{4}\left(\Delta_q^2 + \phi_q^2\right) - 24c\, \phi_Q^2\ , \ \  m_{\sigma_{QQ}{\cal P}^{17}}^2  = 2\lambda_1\Delta_q\phi_Q - 48c\Delta_q\phi_Q\ , \nonumber\\
&& m_{{\cal P}^{17}}^2 = -4\mu_{q}^2 + m_0^2+\lambda_1\left(3\Delta_q^2+\phi_q^2+\phi_Q^2\right) + \frac{\lambda_2}{4}\left(3\Delta_q^2 + \phi_q^2\right) - 24c\, \phi_Q^2 \ .
\end{eqnarray}

The propagator-inverse matrix for $(K_{Qq},\bar{K}_{Qq}, B'_{Qq},\bar{B}'_{Qq})$ in momentum space reads
\begin{eqnarray}
{\cal D}_{KB'}^{-1} \equiv \left(
\begin{array}{cccc}
p^2-m_{{\cal P}^4}^2  & 2i(\mu_Q-\mu_q)p_0 & 0 & -m_{{\cal P}^4{\cal S}^{21}}^2 \\
-2i(\mu_Q-\mu_q)p_0 & p^2-m_{{\cal P}^5}^2 & -m_{{\cal P}^5{\cal S}^{20}}^2 & 0 \\
 0 &  -m_{{\cal P}^5{\cal S}^{20}}^2  & p^2-m_{{\cal S}^{20}}^2 &2i(\mu_Q+\mu_q)p_0 \\
 -m_{{\cal P}^4{\cal S}^{21}}^2  & 0 & -2i(\mu_Q + \mu_q)p_0 & p^2-m_{{\cal S}^{21}}^2 \\
\end{array}
\right)\ ,
\end{eqnarray}
where
\begin{eqnarray}
&& m_{{\cal P}^{4}}^2 = m_{{\cal P}^{5}}^2  = -(\mu_{Q}-\mu_{q})^2 + m_0^2+\lambda_1\left(\Delta_q^2+\phi_q^2+\phi_Q^2\right) + \frac{\lambda_2}{4}\left(\Delta_q^2 + \phi_q^2 - \phi_q\phi_Q + \phi_Q^2\right) - 24c\phi_q\phi_Q\ ,\nonumber\\
&&m_{{\cal S}^{20}}^2 = m_{{\cal S}^{21}}^2 = -(\mu_{Q}+\mu_{q})^2 + m_0^2+\lambda_1\left(\Delta_q^2+\phi_q^2+\phi_Q^2\right) + \frac{\lambda_2}{4}\left(\Delta_q^2 + \phi_q^2 + \phi_q\phi_Q + \phi_Q^2\right) + 24c\phi_q\phi_Q \ , \nonumber\\
&& m_{{\cal P}^4{\cal S}^{21}}^2 = -\frac{\lambda_2}{4}\Delta_q\phi_Q - 24c\Delta_q\phi_Q\ , \ \  m_{{\cal P}^5{\cal S}^{20}}^2 = \frac{\lambda_2}{4}\Delta_q\phi_Q + 24c\Delta_q\phi_Q\ .
\end{eqnarray}

The propagator-inverse matrix for $(\kappa_{Qq},\bar{\kappa}_{Qq}, B_{Qq},\bar{B}_{Qq})$ in momentum space reads
\begin{eqnarray}
{\cal D}_{\kappa B}^{-1} \equiv \left(
\begin{array}{cccc}
p^2-m_{{\cal S}^4}^2  & 2i(\mu_Q-\mu_q)p_0 & 0 & -m_{{\cal S}^4{\cal P}^{21}}^2 \\
-2i(\mu_Q-\mu_q)p_0 & p^2-m_{{\cal S}^5}^2 & -m_{{\cal S}^5{\cal P}^{20}}^2 & 0 \\
 0 &  -m_{{\cal S}^5{\cal P}^{20}}^2  & p^2-m_{{\cal P}^{20}}^2 &2i(\mu_Q+\mu_q)p_0 \\
 -m_{{\cal S}^4{\cal P}^{21}}^2  & 0 & -2i(\mu_Q + \mu_q)p_0 & p^2-m_{{\cal P}^{21}}^2 \\
\end{array}
\right)\ ,
\end{eqnarray}
where
\begin{eqnarray}
&&m_{{\cal S}^{4}}^2 = m_{{\cal S}^{5}}^2  = -(\mu_{Q}-\mu_{q})^2 + m_0^2+\lambda_1\left(\Delta_q^2+\phi_q^2+\phi_Q^2\right) + \frac{\lambda_2}{4}\left(\Delta_q^2 + \phi_q^2 + \phi_q\phi_Q + \phi_Q^2\right) + 24c\phi_q\phi_Q \ ,\nonumber\\
&& m_{{\cal P}^{20}}^2 = m_{{\cal P}^{21}}^2  = -(\mu_{Q}+\mu_{q})^2 + m_0^2+\lambda_1\left(\Delta_q^2+\phi_q^2+\phi_Q^2\right) + \frac{\lambda_2}{4}\left(\Delta_q^2 + \phi_q^2 - \phi_q\phi_Q + \phi_Q^2\right) - 24c\phi_q\phi_Q \ , \nonumber\\
&&m_{ {\cal S}^4{\cal P}^{21}}^2 = -\frac{\lambda_2}{4}\Delta_q\phi_Q - 24c\Delta_q\phi_Q \ , \ \  m_{{\cal S}^5{\cal P}^{20}}^2 = \frac{\lambda_2}{4}\Delta_q\phi_Q + 24c\Delta_q\phi_Q \ .
\end{eqnarray}
\end{widetext}

In the vacuum the light pion mass is evaluated to be
\begin{eqnarray}
({m}_{\pi_{qq}}^{\rm vac})^2 &=& m_0^2+\lambda_1\big[(\phi_{q}^{\rm vac})^2+(\phi_{Q}^{\rm vac})^2\big]  + \frac{\lambda_2}{4}(\phi_{q}^{\rm vac})^2 \nonumber\\
&& - 24c(\phi_{Q}^{\rm vac})^2 \nonumber\\
&=&  \frac{\sqrt{2}\bar{c}m_q}{\phi^{\rm vac}_{q}}\ , \label{MPiVac}
\end{eqnarray}
and the remaining masses read
\begin{eqnarray}
({m}_{\pi_{QQ}}^{\rm vac})^2 &=&  ({m}_{\pi_{qq}}^{\rm vac})^2 + \left(\frac{\lambda_2}{4} + 24c\right)\big[(\phi_{Q}^{\rm vac})^2-(\phi_{q}^{\rm vac})^2\big] \nonumber\\
&=& \frac{\sqrt{2}\bar{c}m_Q}{\phi_Q^{\rm vac}}\ , \label{MPi2Vac}
\end{eqnarray}
\begin{eqnarray}
({m}_{\eta_{qq}}^{\rm vac})^2 = ({m}_{\pi_{qq}}^{\rm vac})^2 + 48c(\phi_{Q}^{\rm vac})^2 \ , \nonumber\\
\end{eqnarray}
\begin{eqnarray}
({m}_{\eta_{QQ}}^{\rm vac})^2 &=& ({m}_{\pi_{qq}}^{\rm vac})^2 + \frac{\lambda_2}{4}\big[(\phi_{Q}^{\rm vac})^2-(\phi_{q}^{\rm vac})^2\big] \nonumber\\
&&  + 24c\big[(\phi_{q}^{\rm vac})^2 + (\phi_{Q}^{\rm vac})^2\big] \ , \end{eqnarray}
\begin{eqnarray}
(m_{\eta_{qq}\eta_{QQ}}^{\rm vac})^2  = 48c\phi^{\rm vac}_{q}\phi^{\rm vac}_{Q} \ ,
\end{eqnarray}
\begin{eqnarray}
({m}_{K_{Qq}}^{\rm vac})^2 &=& ({m}_{\pi_{qq}}^{\rm vac})^2 + \left(\frac{\lambda_2}{4} + 24c\right)\big[(\phi_{Q}^{\rm vac})^2-\phi_{q}^{\rm vac}\phi_{Q}^{\rm vac}\big] \nonumber\\
&=&\frac{\sqrt{2}\bar{c}(m_q+m_Q)}{\phi_{q}^{\rm vac}+\phi_{Q}^{\rm vac}}\ ,
\end{eqnarray}
\begin{eqnarray} 
({m}_{B_{qq}}^{\rm vac})^2 = ({m}_{\pi_{qq}}^{\rm vac})^2 \ ,
\end{eqnarray}
\begin{eqnarray}
({m}_{B_{Qq}}^{\rm vac})^2 =  ({m}_{K_{Qq}}^{\rm vac})^2 , 
\end{eqnarray}
\begin{eqnarray}
({m}_{B_{QQ}}^{\rm vac})^2 = ({m}_{\pi_{QQ}}^{\rm vac})^2 \ ,
\end{eqnarray}
\begin{eqnarray}
({m}_{a_{0,qq}}^{\rm vac})^2 =  ({m}_{\pi_{qq}}^{\rm vac})^2 + \frac{\lambda_2}{2}(\phi_{q}^{\rm vac})^2 + 48c(\phi_{Q}^{\rm vac})^2 \, ,
\end{eqnarray}
\begin{eqnarray}
({m}_{a_{0,QQ}}^{\rm vac})^2 &=& ({m}_{\pi_{qq}}^{\rm vac})^2 + \frac{\lambda_2}{4}\big[3(\phi_{Q}^{\rm vac})^2-(\phi_{q}^{\rm vac})^2\big]  \nonumber\\
&& + 24c\big[(\phi_{q}^{\rm vac})^2+(\phi_{Q}^{\rm vac})^2\big] \ ,
\end{eqnarray}
\begin{eqnarray}
({m}_{\sigma_{qq}}^{\rm vac})^2 =  ({m}_{\pi_{qq}}^{\rm vac})^2 + \left(2\lambda_1+\frac{\lambda_2}{2}\right)(\phi_{q}^{\rm vac})^2 \  ,
\end{eqnarray}
\begin{eqnarray}
({m}_{\sigma_{QQ}}^{\rm vac})^2 &=&  ({m}_{\pi_{qq}}^{\rm vac})^2 + 2\lambda_1(\phi_{Q}^{\rm vac})^2  + \frac{\lambda_2}{4}\big[ 3(\phi_{Q}^{\rm vac})^2  \nonumber\\
&& -(\phi_{q}^{\rm vac})^2\big]   +24c \big[ (\phi_{Q}^{\rm vac})^2-(\phi_{q}^{\rm vac})^2\big] \  , \nonumber\\
\end{eqnarray}
\begin{eqnarray}
({m}_{\sigma_{qq}\sigma_{QQ}}^{\rm vac})^2 &=&  2(\lambda_1 -24c)\phi_{q}^{\rm vac}\phi_{Q}^{\rm vac}\ ,
\end{eqnarray}
\begin{eqnarray}
({m}_{\kappa_{Qq}}^{\rm vac})^2 &=& ({m}_{\pi_{qq}}^{\rm vac})^2 + \left(\frac{\lambda_2}{4} + 24c\right)\big[(\phi_{Q}^{\rm vac})^2+\phi_{q}^{\rm vac}\phi_{Q}^{\rm vac}\big]\ , \nonumber\\
\end{eqnarray}
\begin{eqnarray}
({m}_{B'_{qq}}^{\rm vac})^2 = ({m}_{a_{0,qq}}^{\rm vac})^2  \ ,
\end{eqnarray}
\begin{eqnarray}
({m}_{B'_{Qq}}^{\rm vac})^2 
= ({m}_{\kappa_{Qq}}^{\rm vac})^2\ ,
\end{eqnarray}
and
\begin{eqnarray}
({m}_{B'_{QQ}}^{\rm vac})^2 =({m}_{a_{0,QQ}}^{\rm vac})^2 \ ,\label{MBQQVac}
\end{eqnarray}
Those formulas imply that the hadron masses in the vacuum are independent of the coupling $\lambda_1$ with the present inputs~(\ref{InputPhysical}) except for the $\sigma_{qq}$ - $\sigma_{QQ}$ sector. Mass degeneracies between certain mesons and baryons are due to the original Pauli-G\"{u}rsey $SU(4)$ symmetry.

From Eqs.~(\ref{MPiVac}) and~(\ref{MPi2Vac}), in the vacuum the mass difference of $\big(m^{\rm vac}_{\pi_{QQ}}\big)^2$ and $\big(m^{\rm vac}_{\pi_{qq}}\big)^2$ is evaluated to be
\begin{eqnarray}
\big(m^{\rm vac}_{\pi_{QQ}}\big)^2 - \big(m^{\rm vac}_{\pi_{qq}}\big)^2 &=& \left(\frac{\lambda_2}{4}+24c\right)  \Big[\big(\phi^{\rm vac}_{Q}\big)^2-\big(\phi_{q}^{\rm vac} \big)^2\Big]\, , \nonumber\\ \label{PiMassHierarchy}
\end{eqnarray}
where $\phi^{\rm vac}_{Q}>\phi^{\rm vac}_{q}$ holds with the present inputs. The stability of the potential $V_{\rm LSM}$ in Eq.~(\ref{VLSM}) implies $\lambda_2$ is positive as long as $\lambda_1$ is comparably suppressed as naturally inferred by the large-$N_c$ counting, while $c$ is positive so as to reproduce $m^{\rm vac}_{\eta_{qq}}>m^{\rm vac}_{\pi_{qq}}$ appropriately. Thus, $\lambda_2/4+24c>0$ holds, and our fundamental assumption that $\pi_{QQ}$ is heavier than $\pi_{qq}$ is indeed satisfied from Eq.~(\ref{PiMassHierarchy}).

\section{Hadron mass spectrum with $m_{\pi_{qq}}^{\rm vac}=738$ MeV}
\label{sec:HadronMassHeavyPion}
In the main text we have adopted the physical inputs listed in Eq.~(\ref{InputPhysical}) to present numerical results. Meanwhile, currently QC$_2$D lattice simulations on hadron mass spectra at finite $\mu_q$ have been done with a comparably heavy pion mass for $N_f=2$~\cite{Hands:2007uc,Murakami:2022lmq}. Hence, in this appendix we supplementally exhibit the results with a heavy $m_{\pi_{qq}}^{\rm vac}$.

In particular, following Ref.~\cite{Murakami:2022lmq} here we employ $m_{\pi_{qq}}^{\rm vac}  = 738\, {\rm MeV}$ for pion mass. Besides, the lattice data reported in Ref.~\cite{Iida:2024irv} implies that the pion decay constant would lie in a range of $f_{\pi_{qq}}^{\rm vac}/\sqrt{2} \sim 50$ - $60$ MeV, hence we take $\phi_{q} = f_{\pi_{qq}}^{\rm vac}=\sqrt{2}\times 60$ MeV as a demonstration. As for the heavy sector so far there is no information from lattice simulations. Here we assume  $f_{K_{Qq}}^{\rm vac}/f_{\pi_{qq}}^{\rm vac} =110/92 $ and $m_{K_{Qq}}^{\rm vac}/m_{\pi_{qq}}^{\rm vac} = 494/138$ to determine the kaon decay constant and mass, following the physical values in three-color QCD. On top of them we also take $\lambda_1=0$ and $c=0.15$. With those inputs, one can fix the model parameters as
 \begin{eqnarray}
&& m_0^2 = -(3346\, {\rm MeV})^2\ ,  \ \lambda_1=0\ , \ \lambda_2=6552 \ ,  \ c=0.15\ , \nonumber\\
&& \bar{c}m_q/2 = (254\, {\rm MeV})^3\ , \ \ \bar{c}m_Q/2 = (785\, {\rm MeV})^3\ . \nonumber\\ \label{HeavyParameters}
\end{eqnarray}

\begin{figure*}[t]
\centering
\hspace*{-0.5cm} 
\includegraphics*[scale=0.43]{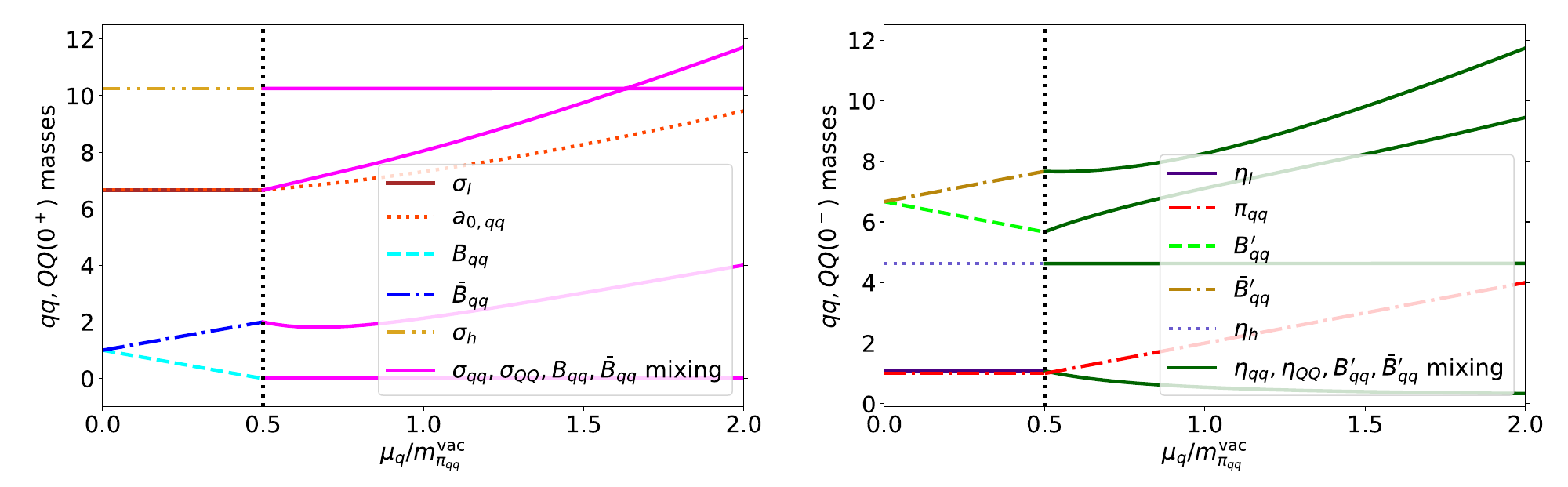}
\caption{Mass spectra of light $0^+$ (left) and $0^-$ (right) hadrons at finite $\mu_q$ ($\mu_q=\mu_Q$) with $m_{\pi_{qq}}^{\rm vac} = 738$ MeV. We have adopted the parameter values in Eq.~(\ref{HeavyParameters}).}
\label{fig:MassLLApp}
\end{figure*}
\begin{figure*}[t]
\centering
\hspace*{-0.5cm} 
\includegraphics*[scale=0.43]{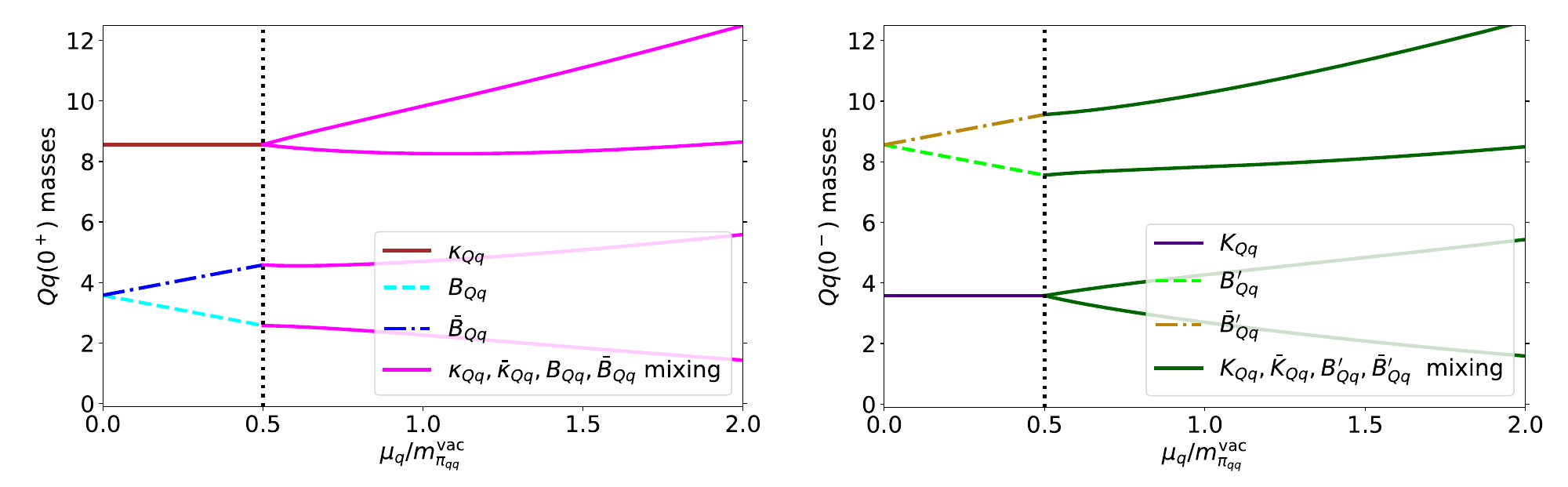}
\caption{Mass spectra of heavy-light $0^+$ (left) and $0^-$ (right) hadrons at finite $\mu_q$ ($\mu_q=\mu_Q$) with $m_{\pi_{qq}}^{\rm vac} = 738$ MeV.}
\label{fig:MassHLApp}
\end{figure*}
\begin{figure*}[t]
\centering
\hspace*{-0.5cm} 
\includegraphics*[scale=0.43]{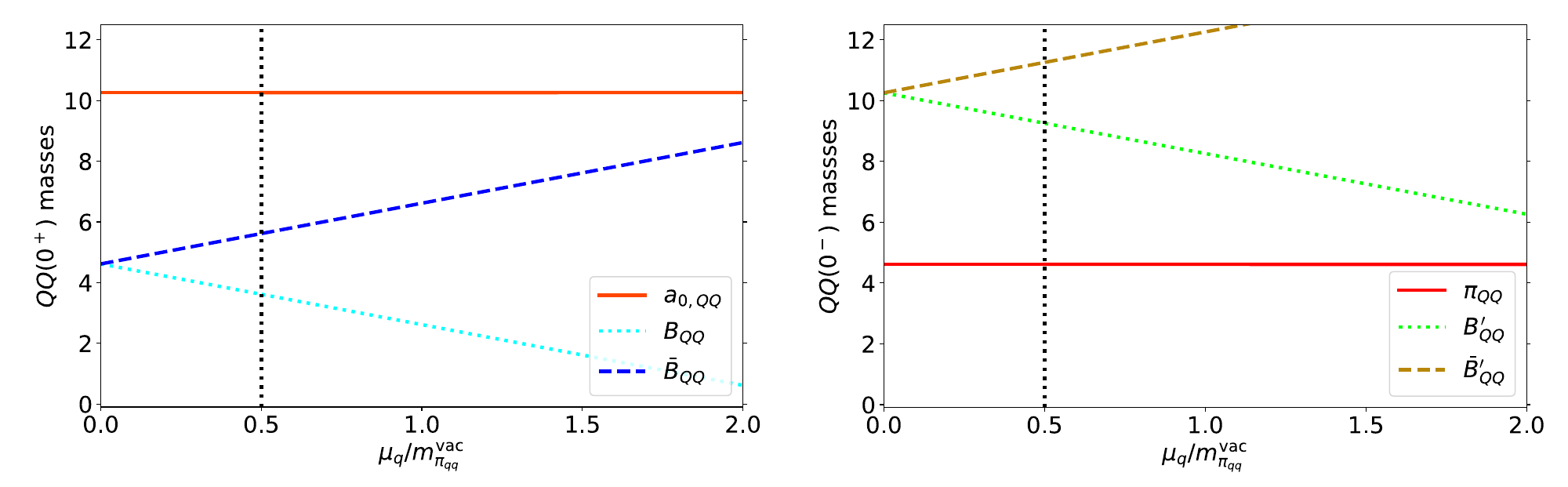}
\caption{Mass spectra of the remaining heavy $0^+$ (left) and $0^-$ (right) hadrons at finite $\mu_q$ ($\mu_q=\mu_Q$) with $m_{\pi_{qq}}^{\rm vac} = 738$ MeV.}
\label{fig:MassHHApp}
\end{figure*}

With these parameters, mass spectra of light, heavy-light, and the remaining heavy hadrons at finite $\mu_q$ ($\mu_q=\mu_Q$) can be plotted as indicated in Figs.~\ref{fig:MassLLApp}, ~\ref{fig:MassHLApp} and~\ref{fig:MassHHApp}, respectively. The detailed behaviors are changed from those in Sec.~\ref{sec:HadronMass}, but qualitatively they agree. It should be noted that a tiny level repulsion is found for the excited $\sigma_{qq}$ - $\sigma_{QQ}$ - $B_{qq}$ - $\bar{B}_{qq}$ mixed state around $\mu_q \approx 1.7 m_{\pi_{qq}}^{\rm vac}$ in Fig.~\ref{fig:MassLLApp}; there is no level crossing.

\section{Detailed analysis of the NJL model}
\label{sec:NJLDetail}

Here we present important properties of the NJL model with $2+2$ flavors and derive the gap equations~(\ref{Gap1NJL}) -~(\ref{Gap3NJL}).

Our NJL model is of the form
\begin{eqnarray}
{\cal L}_{\rm NJL} &=& \bar{\psi}(i\Slash{\partial}+\mu_q\gamma_0-M)\psi+4G{\rm tr}[\Phi^\dagger\Phi] \nonumber\\
&& + \frac{K}{24}\Big( \epsilon_{ijklmnop}\Phi_{ij}\Phi_{kl}\Phi_{mn}\Phi_{op}  + {\rm H.c.}\Big)\, , \label{NJLApp}
\end{eqnarray}
where $\psi=(q_1,q_2,Q_1,Q_2)^T$ is the extended flavor quartet and $M={\rm diag}(m_q,m_q,m_Q,m_Q)$ is a quark mass matrix. The $8\times8$ matrix
\begin{eqnarray}
\Phi_{ij} = \Psi_j^T\sigma^2\tau_c^2\Psi_i
\end{eqnarray}
is a quark composite operator with
\begin{eqnarray}
\Psi &=& (\psi_R,\tilde{\psi}_L)^T \nonumber\\
&=& (q_{1R},q_{2R},Q_{1R},Q_{2R},\tilde{q}_{1L},\tilde{q}_{2L},\tilde{Q}_{1L},\tilde{Q}_{2L})^T\ . \nonumber\\
\end{eqnarray}
The $G$ term is rewritten into ($\tau_f^0 = {\bm 1}_{2\times2}$)
\begin{eqnarray}
{\rm tr}[\Phi^\dagger\Phi] &=&\frac{1}{4}|q^T C\gamma_5\tau_c^2\tau_f^2 q|^2 + \frac{1}{4}|q^T C \tau_c^2\tau_f^2 q|^2 \nonumber\\
&& + \frac{1}{4}\sum_{a=0}^3\big[(\bar{q}\tau_f^aq)^2 + (\bar{q}_1i\gamma_5\tau_f^aq_1)^2\big]   \nonumber\\
&& + (q\leftrightarrow Q)\ ,
\end{eqnarray}
and hence this term is an extended version of the familiar four-point interaction~\cite{Hatsuda:1994pi,Buballa:2003qv}, which preserves $U(8)$ symmetry. Meanwhile, the $K$ term accounts for the $U(1)$ axial anomaly due to its totally asymmetric structure which can be understood as an extension of the KMT determinant coupling. By assuming the following mean fields:
\begin{eqnarray}
&& \langle\bar{q}q\rangle \equiv \langle\bar{q}_1q_1\rangle = \langle\bar{q}_2q_2\rangle \ ,\ \ \langle qq\rangle \equiv  -\frac{1}{2}\langle q^TC\gamma_5\tau_c^2\tau_f^2q\rangle\ , \nonumber\\
&& \langle\bar{Q}Q\rangle \equiv \langle\bar{Q}_1Q_1\rangle = \langle\bar{Q}_2Q_2\rangle \ ,
\end{eqnarray}
and adopting linear approximations
\begin{eqnarray}
XY &\approx&  XY + \langle X\rangle Y+\langle Y\rangle X-\langle X\rangle\langle Y\rangle \ ,
\end{eqnarray}
and
\begin{eqnarray}
WXYZ &\approx& \langle Y\rangle\langle Z\rangle WX +  \langle X\rangle\langle Z\rangle WY +  \langle X\rangle\langle Y\rangle WZ \nonumber\\
&+&   \langle W\rangle\langle Z\rangle XY +  \langle W\rangle\langle Y\rangle XZ +  \langle W\rangle\langle X\rangle YZ \nonumber\\
&+&  \langle X\rangle\langle Y\rangle\langle Z\rangle W + \langle W\rangle\langle Y\rangle \langle Z\rangle X +\langle W\rangle\langle X\rangle\langle Z\rangle Y \nonumber\\
&+&   \langle W\rangle\langle X\rangle\langle Y\rangle Z-3 \langle W\rangle\langle X\rangle\langle Y\rangle\langle Z\rangle \ ,
\end{eqnarray}
the NJL Lagrangian~(\ref{NJLApp}) is reduced to
\begin{eqnarray}
{\cal L}_{\rm NJL} &=&  \bar{\Psi}_q
\left(
\begin{array}{cc}
i\Slash{\partial}+\mu_q\gamma_0-M_q &  -\hat{\Delta} \gamma_5\tau_c^2\tau_f^2 \\
 \hat{\Delta} \gamma_5\tau_c^2\tau_f^2  &  i\Slash{\partial} - \mu_q\gamma_0-M_q\\
\end{array}
\right)\Psi_q  \nonumber\\
&& + \bar{Q}(i\Slash{\partial}+\mu_q\gamma_0-M_Q)Q  \nonumber\\
&&+ \left(G+\frac{K}{2}\sigma_Q^2\right)(\bar{q}i\gamma_5\tau_f^a q)^2  \nonumber\\
&& +(4G+2K\sigma_q\sigma_Q)|\bar{Q}_1i\gamma_5 q_1|^2+ \cdots + {\cal L}_{\rm MF} \  , \nonumber\\
\label{LNJLQC2D}
\end{eqnarray}
where
\begin{eqnarray}
{\cal L}_{\rm MF} &=& -4G\left(\langle qq\rangle^2+\langle\bar{q}q\rangle^2+\langle\bar{Q}Q\rangle^2 \right) \nonumber\\
&& -6K\left(\langle qq\rangle^2+\langle\bar{q}q\rangle^2\right)\langle\bar{Q}Q\rangle^2 
\end{eqnarray}
represents the mean-field contributions. In Eq.~(\ref{LNJLQC2D}),
\begin{eqnarray}
\Psi_q = \frac{1}{\sqrt{2}}(q, q^c)^T\ \ {\rm with}\ \ q^c = C\bar{q}^T
\end{eqnarray}
is the Nambu-Gorkov spinor, and we have defined the following constituent-quark masses and diquark gap:
\begin{eqnarray}
M_q &=& m_q-4G\langle\bar{q}q\rangle - 2K\langle\bar{q}q\rangle\langle\bar{Q}Q\rangle^2\ , \nonumber\\
M_Q &=& m_Q-4G\langle\bar{Q}Q\rangle - 2K\left(\langle qq\rangle^2+\langle\bar{q}q\rangle^2\right)\langle\bar{Q}Q\rangle \ ,\nonumber\\
\hat{\Delta} &=& -2\left(2G + 2K\langle\bar{Q}Q\rangle^2\right)\langle qq\rangle\ .
\end{eqnarray}

From Eq.~(\ref{LNJLQC2D}), the thermodynamics potential (per volume) at $T=0$ within quark one-loop level can be derived to be [$\int_{\bm p} \equiv\int d^3p/(2\pi)^3$]
\begin{eqnarray}
\Omega/V &=& 4G\left(\langle qq\rangle^2+\langle\bar{q}q\rangle^2+\langle\bar{Q}Q\rangle^2 \right) \nonumber\\
&& + 6K\left(\langle qq\rangle^2+\langle\bar{q}q\rangle^2\right)\langle\bar{Q}Q\rangle^2 \nonumber\\
&&  -4\int_{\bm p}\left[ \tilde{\epsilon}_{\rm p}^{(q)}({\bm p}) + \tilde{\epsilon}_{\rm a}^{(q)}({\bm p})\right] \nonumber\\
&& -8\int_{\bm p}\left[ E_{\bm p}^{(Q)} + \theta(\mu_q-E_{\bm p}^{(Q)}) (\mu-E_{\bm p}^{(Q)}) \right] \ , \nonumber\\ \label{PotentialNJL}
\end{eqnarray}
where one-particle excitation energies read ($f=q,Q$)
\begin{eqnarray}
\epsilon_{\rm p}^{(f)}({\bm p}) &=& E^{(f)}_{\bm p}-\mu_q\ ,  \nonumber\\
\epsilon_{\rm a}^{(f)}({\bm p}) &=& E^{(f)}_{\bm p} + \mu_q \ , \nonumber\\
\tilde{\epsilon}_{\rm p}^{(q)}({\bm p}) &=& \sqrt{(E_{\bm p}^{(q)}-\mu_q)^2 + \hat{\Delta}^2}\ ,\nonumber\\
\tilde{\epsilon}_{\rm a}^{(q)}({\bm p}) &=& \sqrt{(E_{\bm p}^{(q)}+\mu_q)^2 + \hat{\Delta}^2}\ ,
\end{eqnarray}
with $E^{(f)}_{\bm p} = \sqrt{{\bm p}^2+M_f^2}$. From the potential~(\ref{PotentialNJL}) gap equations with respect to $\langle\bar{q}q\rangle$, $\langle\bar{Q}Q\rangle$ and $\langle qq\rangle$ are derived, which yields Eqs.~(\ref{Gap1NJL}) -~(\ref{Gap3NJL}). We note that a three-dimensional cutoff $\Lambda$ would be introduced to regularize the divergences.

Hadron masses within the NJL model framework are determined by pole positions of the corresponding amplitude ${\cal T}$ via the Bethe-Salpeter equation ${\cal T} = {\cal K}/(1-{\cal K}{\cal J})$, where ${\cal K}$ and ${\cal J}$ are kernels and loop functions obtained from Eq.~(\ref{LNJLQC2D}). In the hadronic phase ($\langle qq\rangle=0$), those for pions and kaons are derived to be
\begin{eqnarray}
{\cal K}_{\pi_{qq}} &=& -2\left(2G+K\langle\bar{Q}Q\rangle^2 \right) \nonumber\\
{\cal K}_{K_{Qq}} &=& -2\left(2G + K\langle\bar{q}q\rangle\langle\bar{Q}Q\rangle \right) \label{Kernel}
\end{eqnarray}
and
\begin{eqnarray}
{\cal J}_{\pi_{qq}}(q_0) &=& 4\int\frac{d^3p}{(2\pi)^3}\left( \frac{1}{q_0-2E_{\bm p}^{(q)}} -   \frac{1}{q_0+2E^{(q)}_{\bm p}} \right) \ , \nonumber\\
{\cal J}_{{K_{Qq}}}(q_0) &=&  2\int\frac{d^3p}{(2\pi)^3} \left(1+\frac{{\bm p}^2+M_QM_q}{E_{\bm p}^{(Q)}E_{\bm p}^{(q)}}\right) \nonumber\\
&&\times \left(\frac{1}{q_0-E_{\bm p}^{(Q)}-E^{(q)}_{\bm p}} - \frac{1}{q_0+E^{(Q)}_{\bm p} +E_{\bm p}^{(q)}} \right)\, , \nonumber\\ \label{LoopF}
\end{eqnarray}
at the rest frame ${\bm q}={\bm 0}$. Therefore, solving $1-{\cal K}{\cal J}=0$ with respect to $q_0$ with Eqs.~(\ref{Kernel}) and~(\ref{LoopF}), pion and kaon masses are evaluated.

From the gap equations~(\ref{Gap1NJL}) -~(\ref{Gap3NJL}), one can find that the infinitesimal diquark condensates satisfy
\begin{eqnarray}
&& 1 - 4\left(2G+K\langle\bar{Q}Q\rangle^2\right) \nonumber\\
&& \times \int_{\bm p}\left(\frac{1-2\theta(\mu_q-E_{\bm p}^{(q)})}{E_{\bm p}^{(q)}-\mu_q} + \frac{1}{E_{\bm p}^{(q)}+\mu_q}\right) =0\ . \nonumber\\ \label{SuperfluidNJL}
\end{eqnarray}
Meanwhile, from the Pauli-G\"{u}rsey symmetry, in the hadronic phase the kernel for $B_{qq}$ is identical to the pion one
\begin{eqnarray}
{\cal K}_{B_{qq}}= {\cal K}_{\pi_{qq}}\ .
\end{eqnarray}
Its loop function is also given by the pion one with which its argument is modified as $q_0\to q_0+2\mu_q$ and Pauli blocking effects are taken into account in the integrand, due to its baryon number:
\begin{eqnarray}
&& {\cal J}_{B_{qq}}(q_0)  \nonumber\\
&=& 4\int_{\bm p}\left(\frac{1-2\theta(\mu_q-E_{\bm p}^{(q)})}{q_0+2\mu_q-2E_{\bm p}^{(q)}} - \frac{1}{q_0+2\mu_q+2E_{\bm p}^{(q)}}\right)\, . \nonumber\\
\end{eqnarray}
Hence, diquark mass in the hadronic phase is evaluated by $q_0$ satisfying
\begin{eqnarray}
&& 1 + 8\left(2G+K\langle\bar{Q}Q\rangle^2\right) \nonumber\\
&& \times \int_{\bm p}\left(\frac{1-2\theta(\mu_q-E_{\bm p}^{(q)})}{q_0+2\mu_q-2E_{\bm p}^{(q)}} - \frac{1}{q_0+2\mu_q+2E_{\bm p}^{(q)}}\right)  = 0\, . \nonumber\\ \label{DiquarkMass}
\end{eqnarray}
Equations~(\ref{SuperfluidNJL}) and~(\ref{DiquarkMass}) indicate that the onset of the baryon superfluidity is determined at which the diquark mass vanishes. Moreover, from Eqs.~(\ref{Kernel}) and~(\ref{LoopF}) the vacuum pion mass $m_{\pi_{qq}}^{\rm vac}$ satisfies
\begin{eqnarray}
&& 1 - 4\left(2G+K\langle\bar{Q}Q\rangle^2\right) \nonumber\\
&& \times \int_{\bm p}\left(\frac{1}{E_{\bm p}^{(q)}-m_{\pi_{qq}}^{\rm vac}/2} - \frac{1}{E_{\bm p}^{(q)}+m_{\pi_{qq}}^{\rm vac}/2}\right)  = 0\ , \nonumber\\
\end{eqnarray}
then comparing this condition with Eq.~(\ref{SuperfluidNJL}), one can confirm that the superfluidity starts at $\mu_q=m_{\pi_{qq}}^{\rm vac}/2$ under a natural assumption of $(E_{\bm q}^{(q)}>)M_q>m_{\pi_{qq}}^{\rm vac}/2$.

\bibliography{reference}

@article{ParticleDataGroup:2024cfk,
    author = "Navas, S. and others",
    collaboration = "Particle Data Group",
    title = "{Review of particle physics}",
    doi = "10.1103/PhysRevD.110.030001",
    journal = "Phys. Rev. D",
    volume = "110",
    number = "3",
    pages = "030001",
    year = "2024"
}

@article{Aarts:2015tyj,
    author = "Aarts, Gert",
    title = "{Introductory lectures on lattice QCD at nonzero baryon number}",
    eprint = "1512.05145",
    archivePrefix = "arXiv",
    primaryClass = "hep-lat",
    doi = "10.1088/1742-6596/706/2/022004",
    journal = "J. Phys. Conf. Ser.",
    volume = "706",
    number = "2",
    pages = "022004",
    year = "2016"
}

@article{Abuki:2010jq,
    author = "Abuki, Hiroaki and Baym, Gordon and Hatsuda, Tetsuo and Yamamoto, Naoki",
    title = "{The NJL model of dense three-flavor matter with axial anomaly: the low temperature critical point and BEC-BCS diquark crossover}",
    eprint = "1003.0408",
    archivePrefix = "arXiv",
    primaryClass = "hep-ph",
    reportNumber = "TKYNT-10-02",
    doi = "10.1103/PhysRevD.81.125010",
    journal = "Phys. Rev. D",
    volume = "81",
    pages = "125010",
    year = "2010"
}

@article{Acharyya:2024pqj,
    author = "Acharyya, Nirmalendu and Aich, Prasanjit and Bandyopadhyay, Arkajyoti and Vaidya, Sachindeo",
    title = "{Matrix model of two-color one-flavor QCD: The ultrastrong coupling regime}",
    eprint = "2406.06055",
    archivePrefix = "arXiv",
    primaryClass = "hep-th",
    doi = "10.1103/PhysRevD.110.054016",
    journal = "Phys. Rev. D",
    volume = "110",
    number = "5",
    pages = "054016",
    year = "2024"
}

@article{Alford:2007xm,
 archiveprefix        = {arXiv},
 author               = {Alford, Mark G. and Schmitt, Andreas and Rajagopal, Krishna and Sch{\"a}fer, Thomas},
 eprint               = {0709.4635},
 journal              = {Rev. Mod. Phys.},
 pages                = {1455-1515},
 primaryclass         = {hep-ph},
 reportnumber         = {MIT-CTP-3861},
 slaccitation         = {%%CITATION = ARXIV:0709.4635;%%},
 title                = {{Color superconductivity in dense quark matter}},
 url                  = {http://dx.doi.org/10.1103/RevModPhys.80.1455},
 volume               = {80},
 year                 = {2008},
 }

@article{Astrakhantsev:2020tdl,
    author = "Astrakhantsev, N. and Braguta, V. V. and Ilgenfritz, E. M. and Kotov, A. Yu. and Nikolaev, A. A.",
    title = "{Lattice study of thermodynamic properties of dense QC$_2$D}",
    eprint = "2007.07640",
    archivePrefix = "arXiv",
    primaryClass = "hep-lat",
    doi = "10.1103/PhysRevD.102.074507",
    journal = "Phys. Rev. D",
    volume = "102",
    number = "7",
    pages = "074507",
    year = "2020"
}

@article{Baluni:1978rf,
    author = "Baluni, Varouzhan",
    title = "{CP Violating Effects in QCD}",
    reportNumber = "MIT-CTP-726",
    doi = "10.1103/PhysRevD.19.2227",
    journal = "Phys. Rev. D",
    volume = "19",
    pages = "2227--2230",
    year = "1979"
}

@article{Baym:2017whm,
    author = "Baym, Gordon and Hatsuda, Tetsuo and Kojo, Toru and Powell, Philip D. and Song, Yifan and Takatsuka, Tatsuyuki",
    title = "{From hadrons to quarks in neutron stars: a review}",
    eprint = "1707.04966",
    archivePrefix = "arXiv",
    primaryClass = "astro-ph.HE",
    reportNumber = "RIKEN-ITHEMS-REPORT-17, RIKEN-QHP-316, RIKEN-iTHEMS-Report-17",
    doi = "10.1088/1361-6633/aaae14",
    journal = "Rept. Prog. Phys.",
    volume = "81",
    number = "5",
    pages = "056902",
    year = "2018"
}

@article{Boz:2019enj,
    author = "Boz, Tamer and Giudice, Pietro and Hands, Simon and Skullerud, Jon-Ivar",
    title = "{Dense two-color QCD towards continuum and chiral limits}",
    eprint = "1912.10975",
    archivePrefix = "arXiv",
    primaryClass = "hep-lat",
    doi = "10.1103/PhysRevD.101.074506",
    journal = "Phys. Rev. D",
    volume = "101",
    number = "7",
    pages = "074506",
    year = "2020"
}

@article{Braguta:2023yhd,
    author = "Braguta, Victor V.",
    title = "{Phase Diagram of Dense Two-Color QCD at Low Temperatures}",
    doi = "10.3390/sym15071466",
    journal = "Symmetry",
    volume = "15",
    number = "7",
    pages = "1466",
    year = "2023"
}

@article{Brauner:2009gu,
    author = "Brauner, Tomas and Fukushima, Kenji and Hidaka, Yoshimasa",
    title = "{Two-color quark matter: U(1)(A) restoration, superfluidity, and quarkyonic phase}",
    eprint = "0907.4905",
    archivePrefix = "arXiv",
    primaryClass = "hep-ph",
    reportNumber = "YITP-09-44, KUNS-2217",
    doi = "10.1103/PhysRevD.81.119904",
    journal = "Phys. Rev. D",
    volume = "80",
    pages = "074035",
    year = "2009",
    note = "[Erratum: Phys.Rev.D 81, 119904 (2010)]"
}

@article{Buballa:2003qv,
 archiveprefix        = {arXiv},
 author               = {Buballa, Michael},
 eprint               = {hep-ph/0402234},
 journal              = {Phys. Rept.},
 pages                = {205-376},
 primaryclass         = {hep-ph},
 slaccitation         = {%%CITATION = HEP-PH/0402234;%%},
 title                = {{NJL model analysis of quark matter at large density}},
 url                  = {http://dx.doi.org/10.1016/j.physrep.2004.11.004},
 volume               = {407},
 year                 = {2005},
 }

@article{Buividovich:2020dks,
    author = "Buividovich, P. V. and Smith, D. and von Smekal, L.",
    title = "{Electric conductivity in finite-density $SU(2)$ lattice gauge theory with dynamical fermions}",
    eprint = "2007.05639",
    archivePrefix = "arXiv",
    primaryClass = "hep-lat",
    doi = "10.1103/PhysRevD.102.094510",
    journal = "Phys. Rev. D",
    volume = "102",
    number = "9",
    pages = "094510",
    year = "2020"
}

@article{Contant:2019lwf,
    author = "Contant, Romain and Huber, Markus Q.",
    title = "{Dense two-color QCD from Dyson-Schwinger equations}",
    eprint = "1909.12796",
    archivePrefix = "arXiv",
    primaryClass = "hep-ph",
    doi = "10.1103/PhysRevD.101.014016",
    journal = "Phys. Rev. D",
    volume = "101",
    number = "1",
    pages = "014016",
    year = "2020"
}

@article{Duarte:2015ppa,
    author = "Duarte, Dyana C. and Allen, P. G. and Farias, R. L. S. and Manso, Pedro H. A. and Ramos, Rudnei O. and Scoccola, N. N.",
    title = "{BEC-BCS crossover in a cold and magnetized two color NJL model}",
    eprint = "1510.02756",
    archivePrefix = "arXiv",
    primaryClass = "hep-ph",
    doi = "10.1103/PhysRevD.93.025017",
    journal = "Phys. Rev. D",
    volume = "93",
    number = "2",
    pages = "025017",
    year = "2016"
}

@article{Fejos:2025nvd,
    author = "Fejos, G. and Suenaga, D.",
    title = "{Chiral symmetry restoration in QC2D from an effective model using the functional renormalization group}",
    eprint = "2502.10134",
    archivePrefix = "arXiv",
    primaryClass = "hep-ph",
    reportNumber = "RIKEN-iTHEMS-Report-25",
    doi = "10.1103/PhysRevD.111.076018",
    journal = "Phys. Rev. D",
    volume = "111",
    number = "7",
    pages = "076018",
    year = "2025"
}

@article{Fejos:2025oxi,
    author = "Fej{\H{o}}s, Gergely and Suenaga, Daiki",
    title = "{Enhancement of axial anomaly effects in hot two-color QCD: FRG approach in the linear sigma model}",
    eprint = "2506.14010",
    archivePrefix = "arXiv",
    primaryClass = "hep-ph",
    reportNumber = "RIKEN-iTHEMS-Report-25",
    month = "6",
    year = "2025"
}

@article{Fujimoto:2023mzy,
    author = "Fujimoto, Yuki and Kojo, Toru and McLerran, Larry D.",
    title = "{Momentum Shell in Quarkyonic Matter from Explicit Duality: A Dual Model for Cold, Dense QCD}",
    eprint = "2306.04304",
    archivePrefix = "arXiv",
    primaryClass = "nucl-th",
    reportNumber = "INT-PUB-23-018",
    doi = "10.1103/PhysRevLett.132.112701",
    journal = "Phys. Rev. Lett.",
    volume = "132",
    number = "11",
    pages = "112701",
    year = "2024"
}

@article{Fujimoto:2024doc,
    author = "Fujimoto, Yuki and Kojo, Toru and McLerran, Larry",
    title = "{Quarkyonic matter pieces together the hyperon puzzle}",
    eprint = "2410.22758",
    archivePrefix = "arXiv",
    primaryClass = "nucl-th",
    reportNumber = "INT-PUB-24-056, RIKEN-iTHEMS-Report-24",
    month = "10",
    year = "2024"
}

@article{Fukushima:2008su,
    author = "Fukushima, Kenji",
    title = "{Characteristics of the eigenvalue distribution of the Dirac operator in dense two-color QCD}",
    eprint = "0806.1104",
    archivePrefix = "arXiv",
    primaryClass = "hep-ph",
    reportNumber = "YITP-08-45",
    doi = "10.1088/1126-6708/2008/07/083",
    journal = "JHEP",
    volume = "07",
    pages = "083",
    year = "2008"
}

@article{Gursey:1958fzy,
    author = {G\"ursey, Feza},
    title = "{Relation of charge independence and baryon conservation to Pauli\textquoteright{}s transformation}",
    doi = "10.1007/bf02747705",
    journal = "Nuovo Cim.",
    volume = "7",
    number = "3",
    pages = "411--415",
    year = "1958"
}

@article{Hands:2007uc,
    author = "Hands, Simon and Sitch, Peter and Skullerud, Jon-Ivar",
    title = "{Hadron Spectrum in a Two-Colour Baryon-Rich Medium}",
    eprint = "0710.1966",
    archivePrefix = "arXiv",
    primaryClass = "hep-lat",
    doi = "10.1016/j.physletb.2008.01.078",
    journal = "Phys. Lett. B",
    volume = "662",
    pages = "405--412",
    year = "2008"
}

@article{Harada:2010vy,
    author = "Harada, Masayasu and Nonaka, Chiho and Yamaoka, Tetsuro",
    title = "{Masses of vector bosons in two-color dense QCD based on the hidden local symmetry}",
    eprint = "1002.4705",
    archivePrefix = "arXiv",
    primaryClass = "hep-ph",
    doi = "10.1103/PhysRevD.81.096003",
    journal = "Phys. Rev. D",
    volume = "81",
    pages = "096003",
    year = "2010"
}

@article{Harada:2019udr,
    author = "Harada, Masayasu and Liu, Yan-Rui and Oka, Makoto and Suzuki, Kei",
    title = "{Chiral effective theory of diquarks and the $U_A(1)$ anomaly}",
    eprint = "1912.09659",
    archivePrefix = "arXiv",
    primaryClass = "hep-ph",
    doi = "10.1103/PhysRevD.101.054038",
    journal = "Phys. Rev. D",
    volume = "101",
    number = "5",
    pages = "054038",
    year = "2020"
}

@article{Hatsuda:1994pi,
 archiveprefix        = {arXiv},
 author               = {Hatsuda, Tetsuo and Kunihiro, Teiji},
 doi                  = {10.1016/0370-1573(94)90022-1},
 eprint               = {hep-ph/9401310},
 journal              = {Phys. Rept.},
 pages                = {221-367},
 primaryclass         = {hep-ph},
 reportnumber         = {UTHEP-270, RYUTHP-94-1},
 slaccitation         = {%%CITATION = HEP-PH/9401310;%%},
 title                = {{QCD phenomenology based on a chiral effective Lagrangian}},
 volume               = {247},
 year                 = {1994},
 }

@article{Iida:2024irv,
    author = "Iida, Kei and Itou, Etsuko and Murakami, Kotaro and Suenaga, Daiki",
    title = "{Lattice study on finite density QC$_{2}$D towards zero temperature}",
    eprint = "2405.20566",
    archivePrefix = "arXiv",
    primaryClass = "hep-lat",
    reportNumber = "YITP-24-68, RIKEN-iTHEMS-Report-24",
    doi = "10.1007/JHEP10(2024)022",
    journal = "JHEP",
    volume = "10",
    pages = "022",
    year = "2024"
}

@article{Imai:2012hr,
    author = "Imai, Shotaro and Toki, Hiroshi and Weise, Wolfram",
    title = "{Quark-Hadron Matter at Finite Temperature and Density in a Two-Color PNJL model}",
    eprint = "1210.1307",
    archivePrefix = "arXiv",
    primaryClass = "nucl-th",
    doi = "10.1016/j.nuclphysa.2013.06.001",
    journal = "Nucl. Phys. A",
    volume = "913",
    pages = "71--102",
    year = "2013"
}

@article{Kawaguchi:2023olk,
    author = "Kawaguchi, Mamiya and Suenaga, Daiki",
    title = "{Fate of the topological susceptibility in two-color dense QCD}",
    eprint = "2305.18682",
    archivePrefix = "arXiv",
    primaryClass = "hep-ph",
    doi = "10.1007/JHEP08(2023)189",
    journal = "JHEP",
    volume = "08",
    pages = "189",
    year = "2023"
}

@article{Kawaguchi:2024iaw,
    author = "Kawaguchi, Mamiya and Suenaga, Daiki",
    title = "{Sound velocity peak induced by the chiral partner in dense two-color QCD}",
    eprint = "2402.00430",
    archivePrefix = "arXiv",
    primaryClass = "hep-ph",
    doi = "10.1103/PhysRevD.109.096034",
    journal = "Phys. Rev. D",
    volume = "109",
    number = "9",
    pages = "096034",
    year = "2024"
}

@article{Khunjua:2020xws,
    author = "Khunjua, T. G. and Klimenko, K. G. and Zhokhov, R. N.",
    title = "{The dual properties of chiral and isospin asymmetric dense quark matter formed of two-color quarks}",
    eprint = "2003.10562",
    archivePrefix = "arXiv",
    primaryClass = "hep-ph",
    doi = "10.1007/JHEP06(2020)148",
    journal = "JHEP",
    volume = "06",
    pages = "148",
    year = "2020"
}

@article{Kobayashi:1970ji,
    author = "Kobayashi, M. and Maskawa, T.",
    title = "{Chiral symmetry and eta-x mixing}",
    doi = "10.1143/PTP.44.1422",
    journal = "Prog. Theor. Phys.",
    volume = "44",
    pages = "1422--1424",
    year = "1970"
}

@article{Kobayashi:1971qz,
    author = "Kobayashi, M. and Kondo, H. and Maskawa, T.",
    title = "{Symmetry breaking of the chiral u(3) x u(3) and the quark model}",
    doi = "10.1143/PTP.45.1955",
    journal = "Prog. Theor. Phys.",
    volume = "45",
    pages = "1955--1959",
    year = "1971"
}

@article{Kogut:1999iv,
    author = "Kogut, J. B. and Stephanov, Misha A. and Toublan, D.",
    title = "{On two color QCD with baryon chemical potential}",
    eprint = "hep-ph/9906346",
    archivePrefix = "arXiv",
    reportNumber = "ITP-SB-99-28, SUNY-NTG-99-23",
    doi = "10.1016/S0370-2693(99)00971-5",
    journal = "Phys. Lett. B",
    volume = "464",
    pages = "183--191",
    year = "1999"
}

@article{Kogut:2000ek,
    author = "Kogut, J. B. and Stephanov, Misha A. and Toublan, D. and Verbaarschot, J. J. M. and Zhitnitsky, A.",
    title = "{QCD - like theories at finite baryon density}",
    eprint = "hep-ph/0001171",
    archivePrefix = "arXiv",
    reportNumber = "SUNY-NTG-00-11",
    doi = "10.1016/S0550-3213(00)00242-X",
    journal = "Nucl. Phys. B",
    volume = "582",
    pages = "477--513",
    year = "2000"
}

@article{Chiba:2023ftg,
    author = "Chiba, Ryuji and Kojo, Toru",
    title = "{Sound velocity peak and conformality in isospin QCD}",
    eprint = "2304.13920",
    archivePrefix = "arXiv",
    primaryClass = "hep-ph",
    doi = "10.1103/PhysRevD.109.076006",
    journal = "Phys. Rev. D",
    volume = "109",
    number = "7",
    pages = "076006",
    year = "2024"
}

@article{Chiba:2024cny,
    author = "Chiba, Ryuji and Kojo, Toru and Suenaga, Daiki",
    title = "{Thermal effects on sound velocity peak and conformality in isospin QCD}",
    eprint = "2403.02538",
    archivePrefix = "arXiv",
    primaryClass = "hep-ph",
    doi = "10.1103/PhysRevD.110.054037",
    journal = "Phys. Rev. D",
    volume = "110",
    number = "5",
    pages = "054037",
    year = "2024"
}

@article{Kojo:2024sca,
    author = "Kojo, Toru and Suenaga, Daiki and Chiba, Ryuji",
    title = "{Isospin QCD as a Laboratory for Dense QCD}",
    eprint = "2406.11059",
    archivePrefix = "arXiv",
    primaryClass = "hep-ph",
    doi = "10.3390/universe10070293",
    journal = "Universe",
    volume = "10",
    number = "7",
    pages = "293",
    year = "2024"
}

@article{Lenaghan:2001sd,
    author = "Lenaghan, J. T. and Sannino, F. and Splittorff, K.",
    title = "{The Superfluid and conformal phase transitions of two color QCD}",
    eprint = "hep-ph/0107099",
    archivePrefix = "arXiv",
    doi = "10.1103/PhysRevD.65.054002",
    journal = "Phys. Rev. D",
    volume = "65",
    pages = "054002",
    year = "2002"
}

@book{manohar2000heavy,
  title={Heavy Quark Physics},
  author={Manohar, A.V. and Wise, M.B. and Ericson, T. and Landshoff, P.Y.},
  isbn={9780521642415},
  lccn={99038167},
  series={Cambridge Monographs on Particle Physics, Nuclear Physics and Cosmology},
  url={https://books.google.co.jp/books?id=codDQK5OQDIC},
  year={2000},
  publisher={Cambridge University Press}
}

@article{Murakami:2022lmq,
    author = "Murakami, Kotaro and Suenaga, Daiki and Iida, Kei and Itou, Etsuko",
    title = "{Measurement of hadron masses in 2-color finite density QCD}",
    eprint = "2211.13472",
    archivePrefix = "arXiv",
    primaryClass = "hep-lat",
    reportNumber = "YITP-22-146, RIKEN-iTHEMS-Report-22",
    doi = "10.22323/1.430.0154",
    journal = "PoS",
    volume = "LATTICE2022",
    pages = "154",
    year = "2023"
}

@article{Muroya:2003qs,
    author = "Muroya, Shin and Nakamura, Atsushi and Nonaka, Chiho and Takaishi, Tetsuya",
    title = "{Lattice QCD at finite density: An Introductory review}",
    eprint = "hep-lat/0306031",
    archivePrefix = "arXiv",
    doi = "10.1143/PTP.110.615",
    journal = "Prog. Theor. Phys.",
    volume = "110",
    pages = "615--668",
    year = "2003"
}

@article{Nagata:2021ugx,
    author = "Nagata, Keitaro",
    title = "{Finite-density lattice QCD and sign problem: Current status and open problems}",
    eprint = "2108.12423",
    archivePrefix = "arXiv",
    primaryClass = "hep-lat",
    doi = "10.1016/j.ppnp.2022.103991",
    journal = "Prog. Part. Nucl. Phys.",
    volume = "127",
    pages = "103991",
    year = "2022"
}

@article{Pauli:1957voo,
    author = "Pauli, W.",
    title = "{On the conservation of the Lepton charge}",
    doi = "10.1007/bf02827771",
    journal = "Nuovo Cim.",
    volume = "6",
    number = "1",
    pages = "204--215",
    year = "1957"
}

@article{Ratti:2004ra,
    author = "Ratti, Claudia and Weise, Wolfram",
    title = "{Thermodynamics of two-colour QCD and the Nambu Jona-Lasinio model}",
    eprint = "hep-ph/0406159",
    archivePrefix = "arXiv",
    doi = "10.1103/PhysRevD.70.054013",
    journal = "Phys. Rev. D",
    volume = "70",
    pages = "054013",
    year = "2004"
}

@article{Strodthoff:2011tz,
    author = "Strodthoff, Nils and Schaefer, Bernd-Jochen and von Smekal, Lorenz",
    title = "{Quark-meson-diquark model for two-color QCD}",
    eprint = "1112.5401",
    archivePrefix = "arXiv",
    primaryClass = "hep-ph",
    doi = "10.1103/PhysRevD.85.074007",
    journal = "Phys. Rev. D",
    volume = "85",
    pages = "074007",
    year = "2012"
}

@article{Suenaga:2019jjv,
    author = "Suenaga, Daiki and Kojo, Toru",
    title = "{Gluon propagator in two-color dense QCD: Massive Yang-Mills approach at one-loop}",
    eprint = "1905.08751",
    archivePrefix = "arXiv",
    primaryClass = "hep-ph",
    doi = "10.1103/PhysRevD.100.076017",
    journal = "Phys. Rev. D",
    volume = "100",
    number = "7",
    pages = "076017",
    year = "2019"
}

@article{Suenaga:2022uqn,
    author = "Suenaga, Daiki and Murakami, Kotaro and Itou, Etsuko and Iida, Kei",
    title = "{Probing the hadron mass spectrum in dense two-color QCD with the linear sigma model}",
    eprint = "2211.01789",
    archivePrefix = "arXiv",
    primaryClass = "hep-ph",
    reportNumber = "RIKEN-iTHEMS-Report-22, YITP-22-127",
    doi = "10.1103/PhysRevD.107.054001",
    journal = "Phys. Rev. D",
    volume = "107",
    number = "5",
    pages = "054001",
    year = "2023"
}

@article{Suenaga:2023tcy,
    author = "Suenaga, Daiki and Oka, Makoto",
    title = "{Axial anomaly effect on the chiral-partner structure of diquarks at high temperature}",
    eprint = "2305.09730",
    archivePrefix = "arXiv",
    primaryClass = "hep-ph",
    doi = "10.1103/PhysRevD.108.014030",
    journal = "Phys. Rev. D",
    volume = "108",
    number = "1",
    pages = "014030",
    year = "2023"
}

@article{Suenaga:2023xwa,
    author = "Suenaga, Daiki and Murakami, Kotaro and Itou, Etsuko and Iida, Kei",
    title = "{Mass spectrum of spin-one hadrons in dense two-color QCD: Novel predictions by extended linear sigma model}",
    eprint = "2312.17017",
    archivePrefix = "arXiv",
    primaryClass = "hep-ph",
    reportNumber = "YITP-23-172, RIKEN-iTHEMS-Report-23",
    doi = "10.1103/PhysRevD.109.074031",
    journal = "Phys. Rev. D",
    volume = "109",
    number = "7",
    pages = "074031",
    year = "2024"
}

@article{Suenaga:2024vwr,
    author = "Suenaga, Daiki and Oka, Makoto",
    title = "{Fate of {\ensuremath{\Sigma}}c, {\ensuremath{\Xi}}c' and {\ensuremath{\Omega}}c baryons at high temperature with chiral-symmetry restoration}",
    eprint = "2411.12172",
    archivePrefix = "arXiv",
    primaryClass = "hep-ph",
    doi = "10.1103/PhysRevD.111.074032",
    journal = "Phys. Rev. D",
    volume = "111",
    number = "7",
    pages = "074032",
    year = "2025"
}

@article{Sun:2007fc,
    author = "Sun, Gao-feng and He, Lianyi and Zhuang, Pengfei",
    title = "{BEC-BCS crossover in the Nambu-Jona-Lasinio model of QCD}",
    eprint = "hep-ph/0703159",
    archivePrefix = "arXiv",
    doi = "10.1103/PhysRevD.75.096004",
    journal = "Phys. Rev. D",
    volume = "75",
    pages = "096004",
    year = "2007"
}

@article{tHooft:1976snw,
    author = "'t Hooft, Gerard",
    editor = "Shifman, Mikhail A.",
    title = "{Computation of the Quantum Effects Due to a Four-Dimensional Pseudoparticle}",
    reportNumber = "PRINT-76-0551 (HARVARD)",
    doi = "10.1103/PhysRevD.14.3432",
    journal = "Phys. Rev. D",
    volume = "14",
    pages = "3432--3450",
    year = "1976",
    note = "[Erratum: Phys.Rev.D 18, 2199 (1978)]"
}

@article{tHooft:1976rip,
    author = "'t Hooft, Gerard",
    editor = "Shifman, Mikhail A.",
    title = "{Symmetry Breaking Through Bell-Jackiw Anomalies}",
    reportNumber = "PRINT-76-0254 (HARVARD)",
    doi = "10.1103/PhysRevLett.37.8",
    journal = "Phys. Rev. Lett.",
    volume = "37",
    pages = "8--11",
    year = "1976"
}

@article{Witten:1979kh,
    author = "Witten, Edward",
    title = "{Baryons in the 1/n Expansion}",
    reportNumber = "HUTP-79-A007",
    doi = "10.1016/0550-3213(79)90232-3",
    journal = "Nucl. Phys. B",
    volume = "160",
    pages = "57--115",
    year = "1979"
}

@article{Suenaga:2025sln,
    author = "Suenaga, Daiki",
    title = "{Chiral Effective Model of Cold and Dense Two-Color QCD: The Linear Sigma Model Approach}",
    eprint = "2502.04496",
    archivePrefix = "arXiv",
    primaryClass = "hep-ph",
    doi = "10.3390/sym17010124",
    journal = "Symmetry",
    volume = "17",
    number = "1",
    pages = "124",
    year = "2025"
}

@article{Kawaguchi:2020qvg,
    author = "Kawaguchi, Mamiya and Matsuzaki, Shinya and Tomiya, Akio",
    title = "{Analysis of nonperturbative flavor violation at chiral crossover criticality in QCD}",
    eprint = "2005.07003",
    archivePrefix = "arXiv",
    primaryClass = "hep-ph",
    doi = "10.1103/PhysRevD.103.054034",
    journal = "Phys. Rev. D",
    volume = "103",
    number = "5",
    pages = "054034",
    year = "2021"
}

@article{Schafer:1998ef,
    author = {Sch{\"a}fer, Thomas and Wilczek, Frank},
    title = "{Continuity of quark and hadron matter}",
    eprint = "hep-ph/9811473",
    archivePrefix = "arXiv",
    reportNumber = "IASSNS-HEP-98-100",
    doi = "10.1103/PhysRevLett.82.3956",
    journal = "Phys. Rev. Lett.",
    volume = "82",
    pages = "3956--3959",
    year = "1999"
}

\end{document}